\documentclass[longaut,structabstract]{aa} 
\usepackage{graphicx}
\usepackage{color}
\usepackage{txfonts}
\usepackage{longtable}
\usepackage{supertabular}
\usepackage{natbib}
\defcitealias{vanderBurg+14}{vdB14}
\begin{document}

\definecolor{green}{rgb}{0,0.5,0}
\definecolor{grey}{rgb}{0.4,0.5,0.7}
\newcommand{\ab}[1]{\textcolor{red}{\bf Andrea: #1}}
\newcommand{\tbd}[1]{\textcolor{red}{\bf #1}}
\newcommand{\msun}{M_{\odot}}
\newcommand{\ks}{\mathrm{km~s}^{-1}}
\newcommand{\rv}{r_{\Delta}}
\newcommand{\mv}{M_{\Delta}}
\newcommand{\rtwo}{r_{200}}
\newcommand{\vtwo}{{\rm v}_{200}}
\newcommand{\rtwof}{r_{200}^{(0)}}
\newcommand{\rfive}{r_{500}}
\newcommand{\rtwofive}{r_{2500}}
\newcommand{\rume}{r_{\rm{200,U}}}
\newcommand{\mtwo}{M_{200}}
\newcommand{\rhor}{\rho(r)}
\newcommand{\br}{\beta(r)}
\newcommand{\cv}{c_{\Delta}}
\newcommand{\rs}{r_{-2}}
\newcommand{\rh}{r_{\rm H}}
\newcommand{\rb}{r_{\rm B}}
\newcommand{\ri}{r_{\rm I}}
\newcommand{\rc}{r_{\rm c}}
\newcommand{\rn}{r_{\nu}}
\newcommand{\rr}{r_{\rho}}
\newcommand{\ra}{r_{\beta}}
\newcommand{\slos}{\sigma_{\rm{los}}}
\newcommand{\sigv}{\sigma_{\rm{v}}}
\newcommand{\qrr}{Q_r(r)}
\newcommand{\qr}{Q(r)}
\newcommand{\nr}{\nu(r)}
\newcommand{\vrf}{{\rm v}_{{\rm rf}}}

\title{The dynamics of $z \sim 1$ clusters of galaxies from the
  GCLASS survey}

\author{A. Biviano\inst{\ref{ABi}} 
\and R.F.J. van der Burg\inst{\ref{RvdB0}}
\and A. Muzzin\inst{\ref{AMu}} 
\and B. Sartoris\inst{\ref{BSa},\ref{ABi}}
\and G. Wilson\inst{\ref{GWi}}
\and H.K.C. Yee\inst{\ref{Hye}}
}

\offprints{A. Biviano, biviano@oats.inaf.it}

\institute{INAF-Osservatorio Astronomico di Trieste, via G. B. Tiepolo 11, 
I-34131, Trieste, Italy\label{ABi} 
\and
Laboratoire AIM, IRFU/Service d’Astrophysique - CEA/DSM - CNRS, 
Universit\'e Paris Diderot, B\^at. 709, CEA-Saclay,
91191 Gif-sur-Yvette Cedex, France\label{RvdB0} 
\and
Kavli Institute for Cosmology, University of Cambridge, 
Madingley Road, Cambridge, CB3 0HA, UK\label{AMu} 
\and
Universit\`a degli studi di Trieste, Dipartimento di Fisica, 
via G. B. Tiepolo 11, I-34131, Trieste, Italy\label{BSa}
\and
Department of Physics and Astronomy, University of California-Riverside, 
900 University Avenue, Riverside, CA 92521, USA\label{GWi}
\and
Department of Astronomy and Astrophysics, University of Toronto, 
50 St. George Street, Toronto, Ontario, Canada M5S 3H4\label{Hye}
}
 
\date{}

\abstract{The dynamics of clusters of galaxies and its evolution
  provide information on their formation and growth, on the nature of
  dark matter and on the evolution of the baryonic components. Poor
  observational constraints exist so far on the dynamics of clusters
  at redshift $z>0.8$.}{We aim to constrain the internal dynamics of
  clusters of galaxies at redshift $z \sim 1$, namely their mass
  profile $M(r)$, velocity anisotropy profile $\br$, and
  pseudo-phase-space density profiles $\qr$ and $\qrr$, obtained from
  the ratio between the mass density profile and the third power of
  the (total and, respectively, radial) velocity dispersion profiles
  of cluster galaxies.}{We use the spectroscopic and photometric
  data-set of 10 clusters at $0.87<z<1.34$ from the Gemini Cluster
  Astrophysics Spectroscopic Survey (GCLASS). We determine the
  individual cluster masses from their velocity dispersions, then
  stack the clusters in projected phase-space.  We investigate the
  internal dynamics of this stack cluster, using the spatial and
  velocity distribution of its member galaxies.  We determine the
  stack cluster $M(r)$ using the \texttt{MAMPOSSt} method of Mamon et
  al., and its $\br$ by direct inversion of the Jeans equation. The
  procedures used to determine the two aforementioned profiles also
  allow us to determine $\qr$ and $\qrr$.}{Several $M(r)$ models are
  statistically acceptable for the stack cluster (Burkert, Einasto,
  Hernquist, NFW).  The stack cluster total mass concentration, $c
  \equiv \rtwo/\rs=4.0_{-0.6}^{+1.0}$, is in agreement with
  theoretical expectations. The total mass distribution is less
  concentrated than both the cluster stellar-mass and the cluster
  galaxies distributions. The stack cluster $\br$ indicates that
  galaxy orbits are isotropic near the cluster center and become
  increasingly radially elongated with increasing cluster-centric
  distance.  Passive and star-forming galaxies have similar $\br$. The
  observed $\br$ is similar to that of dark matter particles in
  simulated cosmological halos. $\qr$ and $\qrr$ are almost power-law
  relations with slopes similar to those predicted from numerical
  simulations of dark matter halos.}{Comparing our results with those
  obtained for lower-redshift clusters, we conclude that the evolution
  of the concentration-total mass relation and pseudo-phase-space
  density profiles agree with the expectations from $\Lambda$CDM
  cosmological simulations. The fact that $\qr$ and $\qrr$ already
  follow the theoretical expectations in $z \sim 1$ clusters suggest
  these profiles are the result of rapid dynamical relaxation
  processes, such as violent relaxation.  The different concentrations
  of the total and stellar mass distribution, and their subsequent
  evolution, can be explained by merging processes of central galaxies
  leading to the formation of the BCG.  The orbits of passive
    cluster galaxies appear to become more isotropic with time, while
    those of star-forming galaxies do not evolve, presumably because
    star-formation is quenched on a shorter timescale than that required for
    orbital isotropization.  }

\keywords{Galaxies: clusters: general; Galaxies: kinematics and dynamics}

\titlerunning{High-$z$ cluster dynamics}
\authorrunning{A. Biviano et al.}

\maketitle

\section{Introduction}
\label{s:intro}
The total mass density profiles ($\rho(r)$) of cosmological halos are
known to be well described by universal profiles such as the NFW
profile of \citet{NFW96,NFW97} and the profile of \citet[][see also
  \citealt{Navarro+04}]{Einasto65}, at least in the radial range from
10 kpc \citep{Schaller+15} to the virial radius\footnote{The radius
  $\rv$ is the radius of a sphere with mass overdensity $\Delta$ times
  the critical density at the cluster redshift. Throughout this paper
  we refer to the $\Delta=200$ radius as the 'virial radius',
  $\rtwo$. Given the cosmological model, the virial mass $\mtwo$
  follows directly from $\rtwo$ once the cluster redshift is known, $G
  \, \mtwo \equiv \Delta/2 \, H_z^2 \, \rtwo^3$, where $H_z$ is the
  Hubble constant at the mean cluster redshift. The virial velocity is
  related to the virial radius by $\vtwo \equiv \sqrt{\Delta/2} \, H_z \, \rtwo$},
$\rtwo$ \citep{DK14}.  This has been confirmed observationally in the
local Universe, in particular on the scale of clusters of galaxies
\citep[e.g.][]{Carlberg+97-mprof,vanderMarel+00,BG03,Kneib+03,APP05,Biviano+13}.
It is important to characterize the shape of the halo mass density
profiles since they carry information on the nature of the matter
contained in cosmological halos as well as on the way these halos form
and evolve.

In general, the NFW and/or Einasto profiles are thought to result from
an initial, fast assembly of halos
\citep[e.g.][]{HJS99_MN,ElZant08,LC11}, where the dominant dynamical
process are those of chaotic mixing and violent relaxation
\citep{Henon64,LyndenBell67}. These profiles are characterized by a
scale radius and a smoothly changing slope with radius. On the other
hand, the pseudo-phase-space density profile, $\qr \equiv
\rho/\sigma^3$, where $\sigma(r)$ is the total velocity dispersion
profile of dark matter (DM) particles, has a simple power-law
behavior, $\qr \propto r^{-\alpha}$, over a wide radial range
\citep[outside 0.1\% of the virial radius;][]{DelPopolo11} with a
universal value of $\alpha$ for a variety of cosmological halos
\citep[e.g.][]{TN01,DML05,KKH08}.  For this reason, \citet{TN01} have
argued that $\qr$ is a more fundamental dynamical quantity than
$\rho(r)$.  A similar power-law behavior (with a slightly different
slope, $\alpha_r$) also characterizes the related quantity $\qrr
\equiv \rho/\sigma_r^3$ \citep{DML05}, where $\sigma_r$ is the radial
component of $\sigma$.

The origin of the power-law behavior of $\qr$ and $\qrr$ is not
completely understood. It has been shown analytically \citep{DML05}
that it can result from the Jeans equation if a linear relation exists
between the logarithmic slope of $\rho(r)$, $\gamma(r) \equiv d \ln
\rho/d \ln r$, and the velocity anisotropy profile
\begin{equation}
\beta(r) \equiv
1-(\sigma_{\theta}^2+\sigma_{\phi}^2)/(2 \, \sigma_r^2), 
\label{e:beta}
\end{equation}
where $\sigma_r$, $\sigma_{\theta}$, and $\sigma_{\phi}$ are the
radial, and the two tangential components, respectively, of the
velocity dispersion, and where $\sigma_{\theta}=\sigma_{\phi}$ is
usually assumed. A possible explanation of why there should be a
linear relation between $\br$ and $\gamma(r)$ has been proposed by
\citet{Hansen09} in terms of the relative shapes of the radial and
tangential velocity distribution functions of bound particles in a
halo. In summary, it appears that the power-law behavior of $\qr$ and
$\qrr$ can be explained as the result of violent relaxation followed
by dynamical equilibrium in a static gravitational potential.

The value of the power-law radial slope of $\qr$ seems to depend on
the cosmological model in which the halos evolve \citep{KKH08}, as
well as on redshift ($z$) because $\qr$ is related to the halo mass
and the halo growth-rate \citep{LC09}. 
\citet{LC09} 
predict the power-law exponent of
$\qr$ to change by $\sim 10$\% over the range $0<z<1$.  Studying how
$\qr$ evolves with redshift could therefore help us understand the
processes of formation and evolution of cosmological halos.

Currently, little is known observationally about the $\qr$ (and
$\qrr$) of clusters.  One of the quantities entering $\qr$, namely
$\rho(r)$, can be determined by several techniques, e.g. via
gravitational lensing, cluster kinematics, and the emission from the
hot intra-cluster medium. The other quantity, $\sigma(r)$
($\sigma_r(r)$ in the case of $\qrr$), cannot be determined
observationally for DM particles, but it can be derived for cluster
galaxies, assuming that their velocity distribution is the same as that
of DM particles.  This requires a large spectroscopic sample of
cluster members.

\citet{MBM14} determined $\qr$ and $\qrr$ for a $z=0.09$ cluster, and
found it to be consistent with a power-law behavior, although with
more negative exponents than found in numerical simulations
($\alpha=-2.3 \pm 0.1$ vs. $-1.84$ and $\alpha_r=-2.3 \pm 0.2$
vs. $-1.92$). \citet{Biviano+13} determined $\qr$ and $\qrr$ for a
$z=0.44$ cluster from the CLASH-VLT survey \citep{Rosati+14} and found
consistency with the theoretically predicted power-law behavior,
although Fig.~17 in their paper suggests slightly less
negative exponents than the theoretical expectations. It is clearly
impossible to draw conclusions based on only two clusters, but the
observed redshift trend of $\alpha$ appears to go in the opposite sense to
the theoretical expectation by \citet{LC09}.

Strictly related to the determination of $\qr$ and $\qrr$ is the
determination of the velocity anisotropy profile $\br$.
\citet{Mahdavi+99,BK04,HL08,MBM14} have found that in low-redshift ($z
\sim 0.1$) groups and clusters $\br \approx 0$ for
non-emission-line/early-type/red galaxies (hereafter generically
referred to as 'PG' for passive galaxies), while $\br > 0$ for
emission-line/late-type/blue galaxies (hereafter generically referred
to as 'SFG' for star-forming galaxies). In other words,  PG are
characterized by mostly isotropic orbits \citep[see
  also][]{vanderMarel+00,KBM04}, while SFG are characterized by increasingly
radially elongated orbits with distance from the cluster center (which
we refer to as 'radius' hereafter). At higher redshifts, ($z \sim
0.4-0.6$) cluster galaxy orbits appear to be more similar across
different galaxy types \citep{BP09,Biviano+13}, and increasingly
radially elongated with radius. This suggests that the orbits of PG
become more isotropic with time, while those of SFG do not evolve.

More clusters are becoming available for the determination of $M(r),
\br, \qr,$ and $\qrr$, both at low-$z$, from the WINGS \citep{Cava+09}
and its extension OmegaWINGS \citep{Gullieuszik+15}, and at
intermediate-$z$, from the CLASH-VLT survey \citep{Rosati+14}. At
higher $z$ all the analyses so far have been limited to the
determination of $M(r)$, generally modeled with an NFW or an isothermal
sphere profile.  \citet{SC13} collected data for 30 clusters at
$z>0.8$, and derived their concentrations and virial masses from
strong and weak lensing analyses, assuming an NFW $M(r)$. The
compilation of \citet{SC13} include all known $z>0.8$ clusters for
which $M(r)$ has been derived so far, except for ``El Gordo'', whose
mass profile has been determined via gravitational lensing by
\citet{Jee+14}.  The mass profiles of some clusters in the compilation
of \citet{SC13} have been determined in more than one study, generally
by gravitational lensing techniques
\citep{Clowe+00,Jee+05,Jee+05b,Lombardi+05,Jee+06,JT09,Jee+09}, and in
some cases by using hydrostatic equilibrium equations based on X-ray data
\citep{Huo+04,Santos+12} and/or the thermal SZ effect
\citep{Adam+15}. In none of these studies have $\br, \qr,$ or $\qrr$
been determined.

In this paper we use the sample of high-$z$ clusters from GCLASS
\citep{Muzzin+12} to investigate their dynamics. We use the kinematics
of cluster members to determine $M(r)$, $\br$, $\qr$, and $\qrr$ of a
stacked sample of 418 cluster members, belonging to 10 clusters at
$0.87<z<1.34$. The structure of this paper is the following.  In
Sect.~\ref{s:sample} we describe our data-set, the selection of
cluster member galaxies, and the stacking procedure. In the following
Sections, we determine $M(r)$ (Sect.~\ref{s:mprof}), $\br$
(Sect.~\ref{s:beta}), and $\qr, \qrr$ (Sect.~\ref{s:qqr}), of the
stack cluster. We discuss our results in Sect.~\ref{s:disc} and
summarize them and draw our conclusions in Sect.~\ref{s:conc}.
Throughout the paper we adopt $H_0=70$ km~s$^{-1}$~Mpc$^{-1}$,
$\Omega_0=0.3$, $\Omega_\Lambda=0.7$.

\section{The sample}
\label{s:sample}
\subsection{The data-set}
\label{ss:data}
The GCLASS sample consists of 10 rich clusters at $0.87<z<1.34$, and
is fully described in \citet{Muzzin+12}. These clusters were selected
using the red-sequence method \citep{GY00} from the SpARCS survey
\citep{Muzzin+09,Wilson+09}. The photometric catalog is described in
\citet{vanderBurg+13} and it is complete down to a median stellar mass
limit of $1.4 \times 10^{10} \msun$ \citep[][vdB14
  hereafter]{vanderBurg+14}. Spectroscopic coverage was obtained
through observations on the GMOS instruments on the Gemini North and
Gemini South telescopes. Spectroscopic targets were chosen upon
prioritization by three criteria, a) clustercentric distance, b)
observed $z'-3.6 \, \mu$m color, and c) $3.6 \, \mu$m flux, in order
of importance, see \citet{Muzzin+12}.  The number of galaxies with
measured $z$ within the fields of the 10 clusters vary between 49 and
162 \citep{Muzzin+12}, with an average\footnote{Here and throughout
  this paper, we adopt the biweight estimation of central location for
  the average, and the biweight estimation of scale for the dispersion
  of a given statistical set \citep{BFG90}.} of 137. The total number
of cluster members in each field is a subset of these redshifts.  In
Table~\ref{t:clist} we list the cluster names, the total number of
redshifts for galaxies in the field of each cluster, the number of
cluster members (identified as described in Sect.~\ref{ss:members}),
their mean spectroscopic redshift $\overline{z}$ and line-of-sight
velocity dispersion $\slos$, and the clusters $\rtwo$ determined from
$\slos$ as described in Sect.~\ref{ss:members}.
The coordinates of the cluster centers
are given in \citetalias[][Table 1]{vanderBurg+14}, and are based on
the positions of the Brightest Cluster Galaxies (BCG), determined by
\citet{Lidman+12}.

\begin{table}
\centering
\caption{The cluster sample}
\label{t:clist}
\begin{tabular}{lrrrrr}
\hline 
\\[-0.2cm]
SpARCS  &  $N_z$ & $N_m$ & $\overline{z}$ & $\slos$ & $\rtwo$ \\[0.15cm]
number  &        &       &                & [$\ks$] & [kpc] \\[0.15cm]
\hline
\\[-0.2cm]
0034 & 137 & 40 & 0.866 & $609_{-66}^{+75}$   & $888 \pm 110$ \\[0.15cm]
0035 &  49 & 21 & 1.336 & $941_{-137}^{+159}$ & $977 \pm 154$ \\[0.15cm]
0036 & 119 & 48 & 0.869 & $911_{-90}^{+99}$   & $1230 \pm 129$ \\[0.15cm]
0215 & 125 & 46 & 1.004 & $758_{-77}^{+85}$   & $953 \pm 103$ \\[0.15cm]
1047 & 147 & 29 & 0.956 & $680_{-86}^{+98}$   & $926 \pm 138$ \\[0.15cm]
1051 & 145 & 32 & 1.034 & $530_{-65}^{+73}$   & $705 \pm 102$ \\[0.15cm]
1613 & 161 & 83 & 0.872 & $1232_{-93}^{+100}$ & $1663 \pm 130$ \\[0.15cm]
1616 & 162 & 43 & 1.155 & $701_{-73}^{+81}$   & $854 \pm 107$ \\[0.15cm]
1634 & 125 & 48 & 1.177 & $835_{-82}^{+91}$   & $1008 \pm 131$ \\[0.15cm]
1638 & 112 & 38 & 1.195 & $585_{-65}^{+73}$   & $769 \pm 117$ \\[0.15cm]
\hline
\end{tabular}
\tablefoot{$N_z$ is the number of galaxies with $z$ in each cluster
  field \citep[see Table 1 in][]{Muzzin+12}, $N_m$ the number of
  cluster members. The mean redshift $\overline{z}$ and line-of-sight
  velocity dispersion $\sigma_{los}$, are computed on all cluster
  members. The 1 $\sigma$ error on $\overline{z}$ is $<0.001$ for all
  clusters.  The virial radius $\rtwo$ is computed from $\slos$ using
  the iterative procedure of \citet{MBB13}.}
\end{table}

\subsection{Cluster membership}
\label{ss:members}
\citet{Muzzin+12} established the cluster membership of galaxies with
$z$ in a simple way, by requiring member galaxies to lie within 1500
$\ks$ of the mean cluster velocity in rest-frame. Their definition was
sufficiently accurate for their purposes. The study of the internal
cluster dynamics requires however a more accurate membership
determination. We proceed as follows.  First, we split the $z$
distribution of each cluster into groups separated by weighted gaps
with value $\geq 8$ \citep[see][for the definition of 'weighted
  gap']{Beers+91}, and we select the richest of these groups. Then, we
apply three algorithms for the definition of cluster membership,
those of \citet{dHK96}, \citet[]['shifting gapper']{Fadda+96}, and
\citet[]['Clean']{MBB13}. All these use the location of galaxies in
projected phase-space $R, \vrf$, where $R$ is the projected radial
distance from the cluster center, $\vrf \equiv c \,
(z-\overline{z})/(1+\overline{z})$ is the rest-frame velocity and
$\overline{z}$ is the mean cluster redshift.  For the 'shifting
gapper' method we adopt the following parameters: 600 kpc for the bin
size, a minimum of 15 galaxies per bin, and 800 $\ks$ for the
significance of the gap in velocity space \citep[see][for details on
  the meaning of these parameters]{Fadda+96}. If a galaxy passes the
membership criteria of at least three of the four algorithms
considered \citep[including that of][]{Muzzin+12}, it is considered as
a {\em bona fide} cluster member. The resulting number of cluster
members is listed in Table~\ref{t:clist}.

Based on the sample of cluster members, we compute the line-of-sight
velocity dispersion, $\slos$, of each cluster. This is corrected for
the errors in measured $z$, which are typically $\Delta z \sim
0.001$, according to \citet{DdZdT80}.  To obtain an estimate of the
cluster virial radius, $\rtwo$, from
$\slos$ we follow the iterative procedure of \citet{MBB13}. This
assumes an NFW model for the mass distribution, the concentration--mass
relation of \citet{Gao+08}, and the model of \citet{ML05b} for the
velocity anisotropy profile, with the same scale radius as that of the
mass profile \citep[as suggested by][]{MBM10}. Using different
concentration--mass relations and velocity anisotropy profiles within
currently accepted models in the literature does not modify the
resulting $\rtwo$ values in a significant way.

\subsection{Stacking the clusters}
\label{ss:stack}
The number of spectroscopic members is too low in any individual
cluster of our sample to allow determination of its mass profile by
the analysis of kinematics (see Table~\ref{t:clist}). We therefore
stack the individual clusters together, under the assumption that
their mass profiles have a similar shape and differ only by their
normalization.  This is a widely used procedure
\citep[e.g.][]{vanderMarel+00,Rines+03,KBM04} and it relies on the
predicted existence of a universal mass profile for cosmological halos
\citep{NFW97}, and the fact that the concentration of halo mass
profiles is only mildly dependent on their mass
\citep[e.g.][]{Gao+08,MDvdB08,DeBoni+13,GGS16}.  The concentration of
a halo mass profile can be defined as $c \equiv \rtwo/\rs$ where $\rs$
is the radius at which $\gamma=-2$.  In our sample, given the
$\overline{z}$ and $\rtwo$ values that we have determined, we expect a
very narrow range in concentration, $c=2.9$--3.2, following the
$c-\mtwo$ relation of \citet{DeBoni+13}.

The stacking is done in projected phase-space after scaling the
cluster-centric distances by the cluster $\rtwo$, $R_n \equiv R/\rtwo$,
and the line-of-sight rest-frame velocities by the cluster $\vtwo$,
$\rm{v}_n \equiv \vrf/\vtwo$. 

We define the properties of the stack cluster by the weighted average
of the properties characterizing its component clusters, using the
number of cluster members as weights.  The stack cluster has a mean
redshift $\overline{z}=1.02 \pm 0.06$, and a virial radius $\rtwo=1076
\pm 96$ kpc, corresponding to a virial mass $\mtwo=(4.5 \pm 1.2)
\times 10^{14} \, \msun$. Since the stack was built by centering the
clusters on the positions of their BCGs, these 10 galaxies are removed
from the stack in all the following analyses. After removing the BCGs,
the stack cluster contains 418 member galaxies, of which 355 are at
$R_n \leq 1$, 273 are PG, 123 are SFG, and the rest do not have a type
classification\footnote{The type classification is based on the $U-V$
  and $V-J$ rest-frame colors \citepalias{vanderBurg+14}.}.
\begin{figure}
\begin{center}
\begin{minipage}{0.5\textwidth}
\resizebox{\hsize}{!}{\includegraphics{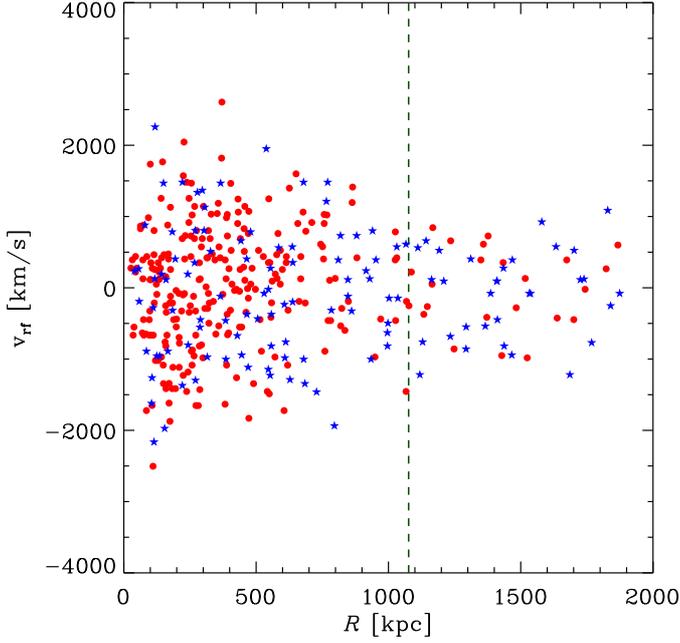}}
\end{minipage}
\end{center}
\caption{The projected phase-space distribution of galaxies in the
  stack cluster, $\vrf$ vs. $R$, obtained by multiplying the
  normalized quantities $\rm{v}_n$ and $R_n$ by the mean values of
  $\vtwo$ and $\rtwo$, resp., for the clusters contributing to the
  stack.  The vertical line indicate the mean value of $\rtwo$. Red
  dots indicate PG, blue stars SFG.}
\label{f:rv}
\end{figure}

\begin{figure}
\begin{center}
\begin{minipage}{0.5\textwidth}
\resizebox{\hsize}{!}{\includegraphics{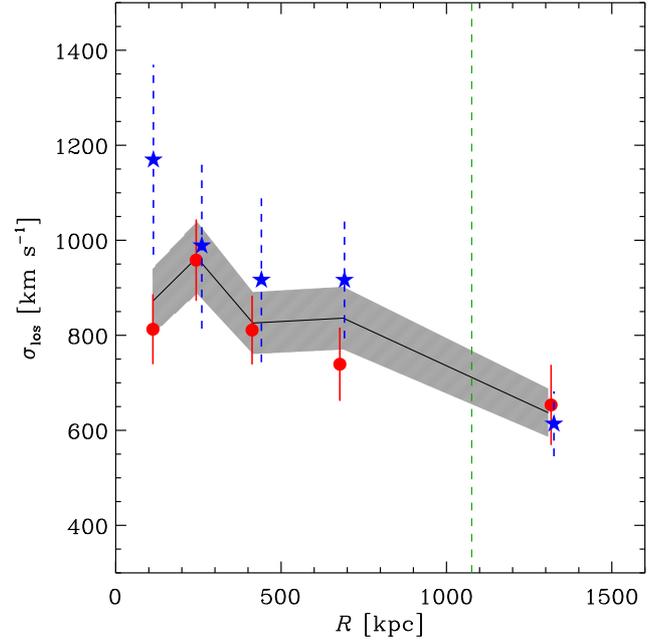}}
\end{minipage}
\end{center}
\caption{The line-of-sight velocity dispersion profile of the stack
  cluster, evaluated on all galaxies (black solid line; the gray
  shading indicates the 1 $\sigma$ confidence level), as well as on PG
  and SFG separately (red dots and blue stars, respectively; error
  bars are 1 $\sigma$)}
\label{f:vdp}
\end{figure}

The projected phase-space distribution of PG and SFG in the stack
cluster is shown in Fig.~\ref{f:rv} \citep[see also][]{Muzzin+14}.
In Fig.~\ref{f:vdp} we show the
line-of-sight velocity dispersion profiles $\slos(R)$ of the stack
cluster, evaluated on all galaxies, and also, separately, on PG and
SFG. It has been shown by \citetalias{vanderBurg+14} that the number
density profiles of PG and SFG in the GCLASS sample are significantly
different.
We confirm this difference by a Kolmogorov-Smirnov test
\citep[e.g.][]{Press+92}, which returns a probability of $<0.001$ that
the radial distributions of PG and SFG are drawn from the same parent
population. On the other hand, the same test applied to the
distributions of rest-frame velocities does not reject the null
hypothesis that the two distributions are drawn from the same parent
one. This is confirmed by the similar $\slos(R)$ of PG and SFG, shown
in Fig.~\ref{f:vdp}. There is however a systematic difference between
the two $\slos(R)$, in that SFG have larger $\slos$ than PG within
$\rtwo$. The ratios of the SFG to the PG $\slos$ is 0.69, 0.97, 0.88,
0.81, and 1.06 in the 5 radial bins shown in Fig.~\ref{f:vdp}. Similar
differences are seen in the $\slos(R)$ of SFG and PG of nearby
\citep[see, e.g., Fig.~13 in][]{Biviano+97} and medium-distant
\citep[see, e.g., Fig.~2 in][]{Carlberg+97-equil} clusters.

\section{The mass profile}
\label{s:mprof}
We use the \texttt{MAMPOSSt} technique \citep{MBB13} to determine the
mass profile of the stack cluster (described in Sect.~\ref{ss:stack}).
\texttt{MAMPOSSt} performs a maximum likelihood analysis of the
distribution of cluster galaxies in projected phase-space, comparing
it with theoretical distributions predicted by the Jeans equation
\citep{BT87} for given models of the mass and velocity anisotropy
profiles. It has been tested on cluster-size halos extracted from
cosmological simulations \citep{MBB13} and applied to several galaxy
clusters already
\citep{Biviano+13,Guennou+14,MBM14,Durret+15,Balestra+16,Pizzuti+16}.

The spatial distribution of galaxies in a stack of the same clusters
used in this paper has already been fitted by
\citetalias{vanderBurg+14}.  We here repeat their analysis to take
into account the fact that our values of the clusters $\rtwo$ are
different from theirs.  We find that the number density profile of all
galaxies members of the stack is best-fitted by an NFW profile with
concentration $c_{\rm{G}}=5.1_{-0.4}^{+0.7}$, a value identical to
that found by \citetalias{vanderBurg+14}.

Having fitted the {\em spatial} distribution of galaxies,
we can restrict the \texttt{MAMPOSSt} maximum likelihood analysis to
the {\em velocity} distribution of cluster galaxies. This is the so-called
'Split' case of the \texttt{MAMPOSSt} procedure \citep{MBB13}.

\begin{table}
\centering
\caption{Results of the \texttt{MAMPOSSt} analysis}
\label{t:mamposst}
\begin{tabular}{llccc}
\hline 
\\[-0.2cm]
$M(r)$ & $\br$ & $\rs$ & Velocity & Likelihood \\
\multicolumn{2}{c}{models} & [Mpc] & Anisotropy & ratio \\
\hline
\\[-0.2cm]
Bur & C  & $0.23_{- 0.04}^{+ 0.17}$ & $1.0_{- 0.1}^{+ 0.6}$ & 0.32 \\[0.15cm]
\bf{Bur} & \bf{O}  & $0.27_{- 0.03}^{+ 0.06}$ & $1.7_{- 0.2}^{+ 2.5}$ & 1.00 \\[0.15cm]
Bur & T  & $0.32_{- 0.03}^{+ 0.12}$ & $1.7_{- 0.1}^{+ 2.4}$ & 0.60 \\[0.15cm]
Ein & C  & $0.28_{- 0.02}^{+ 0.54}$ & $1.0_{- 0.0}^{+ 0.6}$ & 0.10 \\[0.15cm]
Ein & O  & $0.33_{- 0.05}^{+ 0.11}$ & $1.7_{- 0.2}^{+ 2.5}$ & 0.36 \\[0.15cm]
Ein & T  & $0.41_{- 0.05}^{+ 0.29}$ & $1.5_{- 0.1}^{+ 2.1}$ & 0.21 \\[0.15cm]
Her & C  & $0.28_{- 0.03}^{+ 0.41}$ & $1.0_{- 0.1}^{+ 0.6}$ & 0.13 \\[0.15cm]
Her & O  & $0.33_{- 0.04}^{+ 0.08}$ & $2.0_{- 0.2}^{+ 3.1}$ & 0.52 \\[0.15cm]
Her & T  & $0.42_{- 0.05}^{+ 0.20}$ & $1.8_{- 0.1}^{+ 2.7}$ & 0.29 \\[0.15cm]
NFW & C  & $0.26_{- 0.04}^{+ 0.42}$ & $1.0_{- 0.1}^{+ 0.6}$ & 0.13 \\[0.15cm]
NFW & O  & $0.31_{- 0.05}^{+ 0.10}$ & $1.6_{- 0.2}^{+ 2.4}$ & 0.41 \\[0.15cm]
NFW & T  & $0.38_{- 0.04}^{+ 0.25}$ & $1.5_{- 0.1}^{+ 2.1}$ & 0.24 \\[0.15cm]
\hline
\end{tabular}
\tablefoot{The values of $\rs$ for the Her and Bur models have been
  computed from the respective scale radii $r_H$ and $r_B$ using the
  factors 1/2 and 3/2, resp.  The Velocity Anisotropy parameter is
  $\sigma_r/\sigma_{\theta}$, constant at all radii for the C model,
  and evaluated at $r \rightarrow \infty$ for the O and T models. The
  Likelihood ratio is computed relative to the maximum among all
  models. The best-fit model is emphasized in boldface.}
\end{table}

\begin{figure}
\begin{center}
\begin{minipage}{0.5\textwidth}
\resizebox{\hsize}{!}{\includegraphics{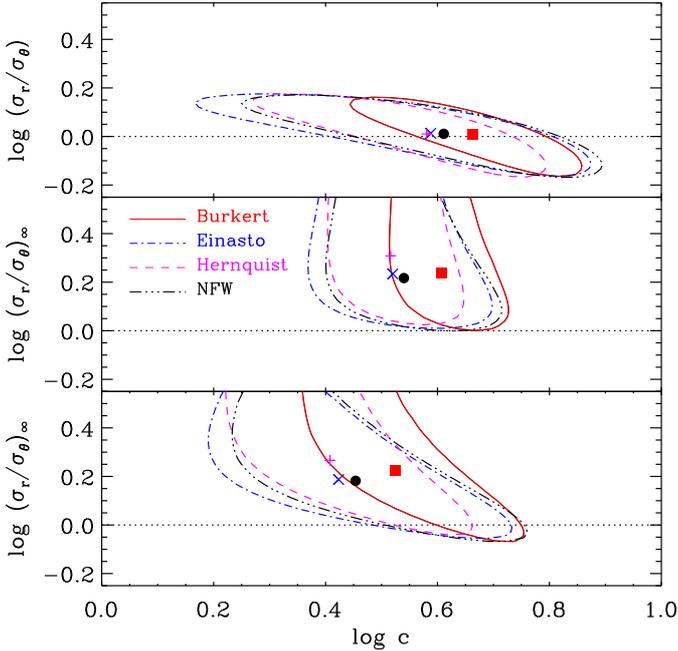}}
\end{minipage}
\end{center}
\caption{The velocity anisotropy parameter $\sigma_r/\sigma_{\theta}$
  at $r \rightarrow \infty$ for the T model (lower panel) and O model
  (middle panel), and constant at all radii for the C model (upper
  panel), vs. the concentration parameter $c$. The best-fit
  \texttt{MAMPOSSt} solutions are indicated by the square, X, plus,
  and dot symbols, and their 1 $\sigma$ confidence levels by the solid
  (red), dash-dotted (blue), dashed magenta, and triple-dot-dashed
  black contours, for the Bur, Ein, Her, and NFW models,
  respectively.}
\label{f:mamposst}
\end{figure}

We consider the four following models for $M(r)$:
\begin{enumerate}
\item the NFW model \citep{NFW97},
\begin{equation}
M(r)=\mtwo {\ln(1+r/\rs)-r/\rs \, (1+r/\rs)^{-1} \over \ln(1+c)-c/(1+c)}
\ ,
\label{e:nfw}
\end{equation}
where $c \equiv \rtwo/\rs$,
characterized by $\gamma(0)=-1$ and $\gamma(\infty)=-3$ \ ,
\item the model of \citet{Einasto65},
\begin{equation}
M(r)=\mtwo {P[3m,2m \, (r/\rs)^{1/m}] \over P[3m,2m \, (\rtwo/\rs)^{1/m}]} \ ,
\label{e:ein}
\end{equation}
'Ein' model hereafter,
where $P(a,x)$ is the regularized incomplete
gamma function. We fix $m=5$, a typical value for
cluster-size halos extracted from cosmological numerical simulations
\citep{MBM10}. The model is characterized by a slope approaching zero near
the center, $\gamma(r)=-2 (r/\rs)^{1/m}$.
\item The model of \citet{Hernquist90},
\begin{equation}
M(r)={\mtwo \, (\rh+\rtwo)^2 \over \rtwo^2} {r^2 \over (r+\rh)^2}
\ ,
\label{e:her}
\end{equation}
'Her' model hereafter, where $\rh=2 \, \rs$. It is characterized
by $\gamma(0)=-1$, like the NFW model, and by a steeper asymptotic slope
$\gamma(\infty)=-4$.
\item The model of \citet{Burkert95},
\begin{eqnarray}
M(r) = \mtwo \, \{ \ln [1+(r/\rb)^2] + 2 \ln (1+r/\rb) \nonumber \\
- 2 \arctan (r/\rb) \} \times \{\ln [1+(\rtwo/\rb)^2] \nonumber \\
+ 2 \ln (1+\rtwo/\rb) - 2 \arctan (\rtwo/\rb) \}^{-1} \ ,
\label{e:bur}
\end{eqnarray} 
'Bur' model hereafter, where $\rb \simeq 2/3 \, \rs$. The model is
characterized by its central core, $\gamma(0)=0$ and an asymptotic
slope like NFW, $\gamma(\infty)=-3$.
\end{enumerate}
All $M(r)$ models are characterized by just two parameters, $\rtwo$
and a scale radius, namely $\rs$ for the Ein and NFW models,
$\rh$ and $\rb$ for the Her and Bur models, respectively. However,
the stack cluster has a fixed value of $\rtwo$ by construction
(see Sect.~\ref{ss:stack}), and so the only remaining free $M(r)$ 
parameter is the scale radius.

As for the velocity anisotropy profiles, we consider three models. The
first, named 'C', is $\br=C \leq 1$, i.e. the velocity anisotropy is
constant at all radii. The second, named 'O' (for opposite) is
characterized by opposite values of the velocity anisotropy at the
cluster center and at very large radii, and it has been introduced by
\citet{Biviano+13},
\begin{equation}
\br=\beta_{\infty} \, {r-\rs \over r+\rs} \ ,
\end{equation}
with $-1 \leq \beta_{\infty} \leq 1$.  The third, named 'T',
is a simplified version of the model of \citet{Tiret+07}, that has
been shown to fit rather well the $\br$ of cluster-sized halos
extracted from numerical simulations \citep{MBM10,MBB13},
\begin{equation}
\br=\beta_{\infty} \, {r \over r+\rs} \ ,
\end{equation}
with $\beta_{\infty} \leq 1$.
In the T model, $\br$ grows from central isotropy to increasingly
radial velocity anisotropy at larger radii. It coincides with
the model of \citet{ML05b} when $\beta_{\infty}=0.5$ and the scale
radius of the \citet{ML05b} model coincides with $\rs$.

All three $\br$ models contribute only one additional free parameter
to the \texttt{MAMPOSSt} analysis, i.e. $C$ or $\beta_{\infty}$, since
$\rs$ in the O and T models is forced to be the same parameter of
$M(r)$.

We run \texttt{MAMPOSSt} on the 355 members of the stack cluster at
$R \leq \rtwo$.  The results of the \texttt{MAMPOSSt} analysis are
given in Table~\ref{t:mamposst}. The errors on each of the two free
parameters of the analysis, $\rs$ and the velocity anisotropy
parameter, are obtained after marginalizing on the other
parameter. Note that instead of listing the values of the $C$ and
$\beta_{\infty}$ velocity anisotropy parameters we list the values of
the related parameters $\sigma_r/\sigma_{\theta}$ and
$(\sigma_r/\sigma_{\theta})_{\infty}$, respectively (note that
$\beta>0$ corresponds to $\sigma_r/\sigma_{\theta}>1$, see
eq.(\ref{e:beta})).  In Fig.~\ref{f:mamposst}, we show the best-fit
results of \texttt{MAMPOSSt} within their 1 $\sigma$ confidence
contours in the plane of the velocity anisotropy parameter
($\sigma_r/\sigma_{\theta}$, constant at all radii for the C model,
and evaluated at $r \rightarrow \infty$ for the O and T models) vs. $c
\equiv \rtwo/\rs$. The value of $\rtwo$ is the same for all models,
being fixed at the value obtained for the stack cluster in
Sect.~\ref{ss:stack}. Both in the Table and in the Figure we use the
factors of 1/2 and 3/2 to convert the scale radii of the Her, and,
respectively, the Bur $M(r)$ to $\rs$ values.

In Table~\ref{t:mamposst} we also list the likelihood ratio with
respect to the maximum obtained among the twelve model
combinations. Formally, the best-fit model combination is Bur+O, but
all other combinations are statistically acceptable.  This means that
we are unable to distinguish the quality of different model fits with
the present data-set. This is also clear from Fig.~\ref{f:mamposst},
where the best-fit solutions for the different models are all within 1
$\sigma$.  C models have however systematically lower likelihoods than
O and T models, and best-fit values of $\beta_{\infty}$ are always positive,
suggesting that a radially increasing velocity anisotropy is
a better fit to the data than a constant velocity anisotropy.

\begin{figure}
\begin{center}
\begin{minipage}{0.5\textwidth}
\resizebox{\hsize}{!}{\includegraphics{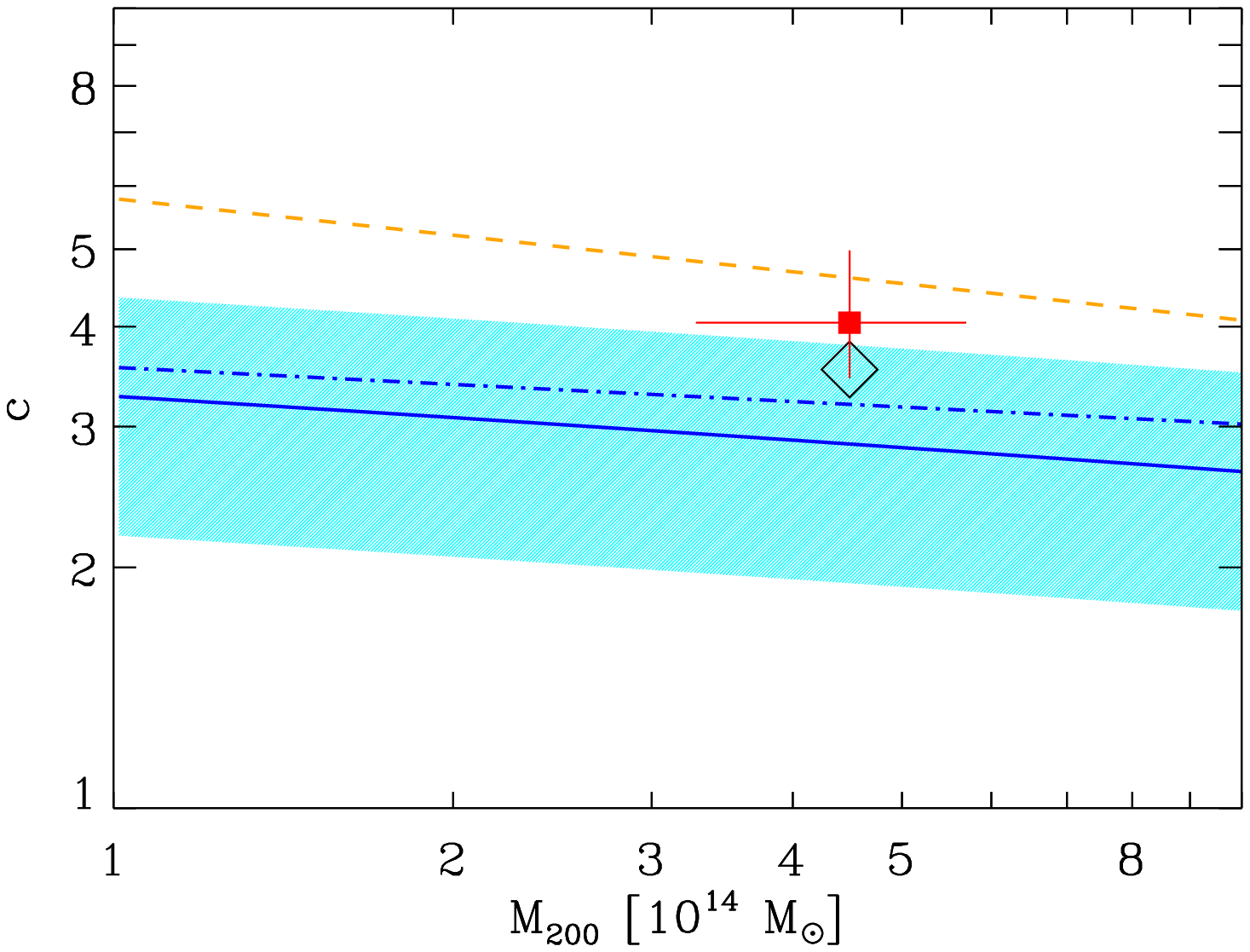}}
\end{minipage}
\end{center}
\caption{The $c$ vs. $\mtwo$ theoretical relations and scatter of
  \citet[][blue solid line and cyan region]{BHHV13} and
  \citet[][blue dash-dotted line]{DeBoni+13}, the observational
  relation of \citet[][orange dashed line]{GGS16}, and the $[c,\mtwo]$
  value obtained for our stack cluster. The $\mtwo$ value is taken
  from Sect.~\ref{ss:stack}; the $c$ value is derived from the $\rtwo$
  value of Sect.~\ref{ss:stack} and either the $\rs$ value of the
  best-fit \texttt{MAMPOSSt} model (Bur+O; red square and 1 $\sigma$
  uncertainties) or that obtained by taking the average of all twelve
  models, weighted according to their relative likelihoods (black
  diamond; the vertical size of the diamond represents the weighted
  dispersion of best-fit concentration parameters, but does not
  include the statistical uncertainty on the measurements).}
\label{f:cM}
\end{figure}

We estimate the weighted average of all $\rs$ values, using the
relative likelihoods of the different models as weights.  We find
$\rs=0.31 \pm 0.02$ Mpc, where the error represents the weighted
dispersion of the different model values. The uncertainties on the
$\rs$ values of the individual models (see Table~\ref{t:mamposst}) are
much larger than the dispersion in the best-fit values among different
models (see also Fig.~\ref{f:mamposst}). In Fig.~\ref{f:cM} we display
the $c$ vs. $\mtwo$ theoretical relations of \citet[][within its
  theoretically predicted scatter]{BHHV13} and \citet{DeBoni+13}, and
the $[c,\mtwo]$ value obtained for our stack cluster. We also display
the phenomenological relation of \citet{GGS16}, which is based on 293
clusters and a variety of observational techniques (lensing, X-ray,
kinematics). All relations are evaluated at the redshift of our stack
cluster, $z=1.02$.  The $\mtwo$ value is taken from
Sect.~\ref{ss:stack}; the $c$ value is derived from the $\rtwo$ value
of Sect.~\ref{ss:stack} and either the $\rs$ value of the best-fit
\texttt{MAMPOSSt} model (Bur+O; $c=4.0_{-0.6}^{+1.0}$, indicated by
the red dot) or that obtained by taking the average of the $\rs$ of
all twelve models, weighted according to their relative likelihoods
($c=3.5 \pm 0.2$, indicated by the black diamond). The uncertainties
in the values of $\rtwo$ and $\rs$ have been propagated to the $c$
uncertainty. We find that the stack cluster has a concentration in
agreement within 1 $\sigma$ with the observational relation of
\citet{GGS16}, and also close to both the theoretical relations of
\citet{DeBoni+13} and \citet{BHHV13}, especially when considering the
large theoretically expected scatter.

\section{The velocity anisotropy profile}
\label{s:beta}
The \texttt{MAMPOSSt} analysis has already provided us with best-fit
values for the parameters of three $\br$ models. Here we derive $\br$
in a non-parametric way, without assuming a model shape for $\br$, by
direct inversion of the Jeans equation, a problem first solved by
\citet{BM82}. In our analysis we follow the procedures of
\citet{SSS90} and \citet[][the latter serving as a check for the
  former in our analysis]{DM92}. We use the same number density
profile parametrization used for the \texttt{MAMPOSSt} procedure in
Sect.~\ref{s:mprof}. We then derive a binned line-of-sight velocity
dispersion profile (see Fig.~\ref{f:vdp}) and smooth it with the
\texttt{LOWESS} technique \citep[see, e.g.,][]{Gebhardt+94}, linearly
extrapolated to infinity (in practice, 30 Mpc from the cluster center)
to a value of 20\% its peak value (we checked that the details of this
extrapolation are irrelevant for our solution within $\rtwo$). For the
Jeans equation inversion we adopt the maximum-likelihood $M(r)$ from
\texttt{MAMPOSSt} (the Bur model of the second line of
Table~\ref{t:mamposst}).

\begin{figure}
\begin{center}
\begin{minipage}{0.5\textwidth}
\resizebox{\hsize}{!}{\includegraphics{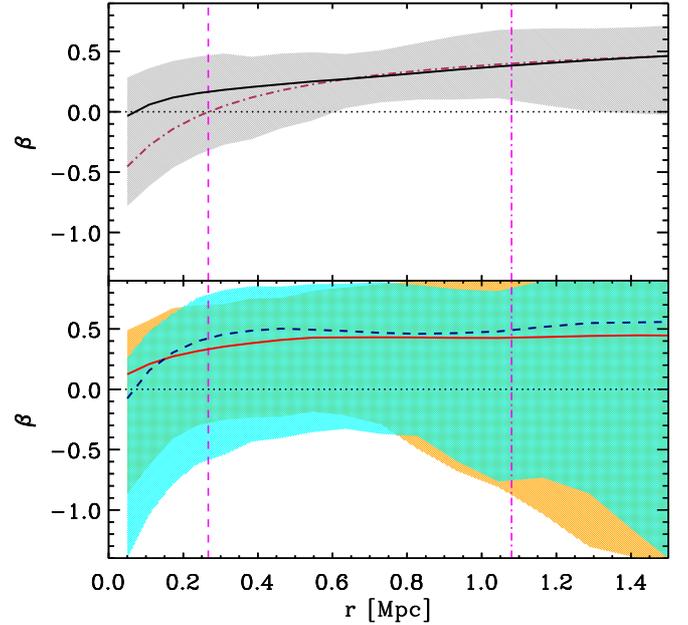}}
\end{minipage}
\end{center}
\caption{The velocity anisotropy profiles $\br$, of different cluster
  galaxy populations as obtained from the inversion of the Jeans
  equation adopting the best-fit \texttt{MAMPOSSt} solution for the
  mass profile.  {\it Top panel:} all cluster members.  The $\br$ is
  represented by the solid (black) curve and the hatched gray region
  represents 68\% confidence levels on the solution.  The best-fit
  \texttt{MAMPOSSt} $\br$ model is represented by the dot-dashed
  (brown) curve.  In both panels, the vertical dashed and dash-dotted
  (magenta) lines indicate the location of $\rs$ and $\rtwo$,
  respectively, and the horizontal dotted line indicates
  $\beta=0$. Below (resp. above) this line, galaxy orbits are more
  tangentially (resp. radially) elongated.  {\it Bottom panel:}
  passive and SF cluster members, separately. Their $\br$ are
  represented by red solid and blue dashed curves, resp. for passive
  and SF galaxies, within their 68\% confidence levels (orange and
  cyan hatched regions, for passive and SF, resp.; green in the
  overlap region).}
\label{f:beta}
\end{figure}

To determine the uncertainties in $\br$ we run 50 bootstrap
resamplings of the galaxies that define the velocity dispersion
profile in the stack cluster, and for each of these resamplings we run
20 random choices of the value of the \texttt{LOWESS} smoothing
parameter. The uncertainties of $\br$ are estimated as the upper and
lower quartiles of the $50 \times 20$ $\br$ bootstrap determinations
with respect to the median. We find that the uncertainties in $\br$
are dominated by the statistics, and not by the systematics in the
smoothing procedure.

In Fig.~\ref{f:beta} (top panel) we show the $\br$ solution and its
68\% confidence levels.  Galaxy orbits are close to isotropic near the
center and become more radially elongated at larger cluster-centric
distances.  Using the best-fit solutions for other $M(r)$ models (Ein,
Her, NFW) (see Table~\ref{t:mamposst}) in the Jeans equation inversion,
does not change $\br$ in a significant way (i.e. the profiles remain
well within the 68\% confidence region shown in Fig.~\ref{f:beta}),
independently on which $\br$ model was assumed in the
\texttt{MAMPOSSt} analysis.

In Fig.~\ref{f:beta} (top panel) we also display the
maximum-likelihood O $\br$ solution found by \texttt{MAMPOSSt} (second
line of Table~\ref{t:mamposst}).  This is in good agreement with the
solution obtained by inversion of the Jeans equation, indicating that
the adopted O model is a good one. On the other hand, the C model
(albeit still acceptable) cannot reproduce the increasing anisotropy
with radius, and this is consistent with the fact that the likelihoods
obtained by \texttt{MAMPOSSt} are systematically lower for the C
models than for the O (and T) models.

We also consider PG and SFG separately, by adopting the same best-fit
Bur $M(r)$ model from \texttt{MAMPOSSt} (since the cluster
gravitational potential is the same for any tracer), but different
number density and velocity dispersion profiles.  The number density
profiles have been determined separately for the two classes of
galaxies, by following the procedure of \citetalias{vanderBurg+14},
and are best-fitted by NFW models with concentrations
$c_{PG}=8.1_{-1.1}^{+0.9}$ and $c_{SFG}=1.5_{-0.4}^{+0.5}$,
respectively \citepalias[these values are very close to those found
  by][]{vanderBurg+14}.  The velocity dispersion profiles for the two
galaxy populations have already been derived in Sect.~\ref{ss:stack}
and are displayed in Fig.~\ref{f:vdp}.  The bottom panel of
Fig.~\ref{f:beta} shows that PG and SFG are characterized by very
similar $\br$, and hence, orbits, within the stack cluster.

\section{The pseudo-phase-space density profiles}
\label{s:qqr}
\begin{figure}
\begin{center}
\begin{minipage}{0.5\textwidth}
\resizebox{\hsize}{!}{\includegraphics{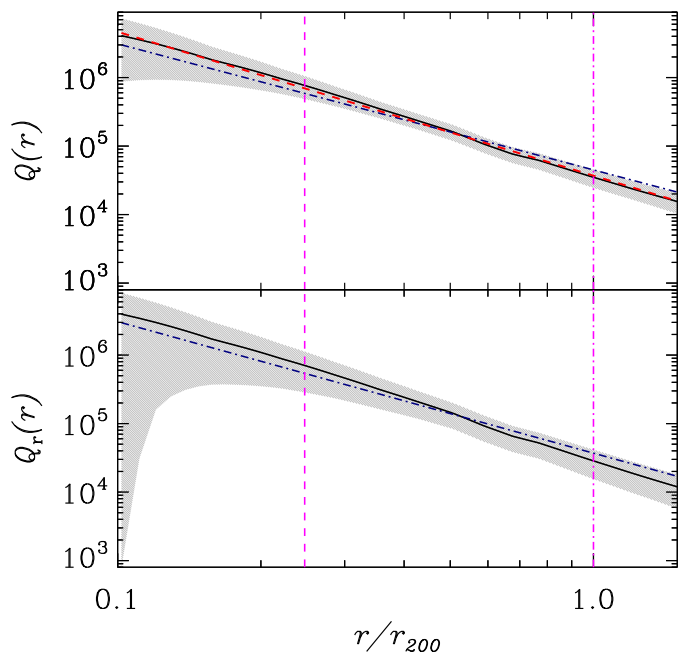}}
\end{minipage}
\end{center}
\caption{Solid lines: the pseudo-phase-space density profiles $\qr
  \equiv \rho/\sigma^3$ (top panel) and $\qrr \equiv
  \rho/\sigma_{\rm{r}}^3$ (bottom panel), as a function of
  cluster-centric radius $r/\rtwo$, within 1 $\sigma$ confidence
  regions (shaded gray regions). The dash-dotted (navy) lines are the
  theoretical relations $\qr \propto r^{-1.84}$ and $\qrr \propto
  r^{-1.92}$ from \citet{DML05}, with free normalization fitted to the
  data. The dashed (red) line in the top panel is the theoretical
  relation $\qr \propto r^{-2.1}$ for massive halos at $z=1$ from
  Fig.~6 in \citet{LC09}. The line is almost indistinguishable from
  the observational relation (solid black line).  In both panels, the
  vertical dashed and dash-dotted (magenta) lines indicate the
  location of $\rs$ and $\rtwo$, respectively.}
\label{f:qqr}
\end{figure}

Using the determination of $M(r)$ (and therefore $\rho(r)$) from
\texttt{MAMPOSSt} (in Sect.~\ref{s:mprof}) and of $\br$ \citep[and
  therefore $\sigma(r)$ and $\sigma_r(r)$, see eqs. 20 and 21
  in][]{SSS90} from the direct inversion of the Jeans equation (in
Sect.~\ref{s:beta}), we determine $\qr$ and $\qrr$. Their
uncertainties are derived from the propagation of the errors on
$\rho(r), \sigma(r),$ and $\sigma_r(r)$. The errors on $\rho(r)$ comes
from the \texttt{MAMPOSSt} analysis, while those on $\sigma(r)$ and
$\sigma_r(r)$ are a byproduct of the bootstrap procedure used to
determine the errors on $\br$. 

In Fig.~\ref{f:qqr} we show the resulting profiles and their 68\%
confidence regions. These profiles are based on the maximum-likelihood
\texttt{MAMPOSSt} model (second line of Table~\ref{t:mamposst}). Other
MAMPOSSt best-fit results based on the T model $\br$ and Ein, Her, or
NFW $M(r)$ give very similar $\qr$ and $\qrr$, well within the
uncertainties, and are not displayed in Fig.~\ref{f:qqr} for the sake
of clarity. We have not considered the C $\br$ models, as these were
shown to be less likely (see Sect.~\ref{s:mprof} and \ref{s:beta}).

In the same figure we also display the theoretical power-law relations
$\qr \propto r^{\alpha}$ and $\qrr \propto r^{\alpha_r}$ with
$\alpha=-1.84, \alpha_r=-1.92$ from \citet{DML05}, based on halos from
cosmological, DM-only simulations, as well as the relation with
$\alpha=-2.1$, valid for massive halos at $z=1$ \citep[see Fig.~6
  in][]{LC09}.  The normalizations of these relations are fitted to
the data in the figure. The observed $\qr$ and $\qrr$ profiles are
very close to the theoretical power-law relations and consistent with
them within the 1 $\sigma$ uncertainties over the full radial range
out to $\rtwo$.

\section{Discussion}
\label{s:disc}
We have investigated the dynamics of a stack of 10 rich clusters from
GCLASS \citep{Muzzin+12}, at mean redshift $\overline{z}=1.02$ and
with mean (inferred from $\slos$) mass $\mtwo=(4.5 \pm 1.2) \, 10^{14} \,
\msun$. To infer the mass and velocity-anisotropy profiles, $M(r)$ and
$\beta(r)$, we have applied the \texttt{MAMPOSSt} technique of
\citet{MBB13}, and the Jeans inversion procedure of \citet{SSS90}.
Hereafter we discuss our results for $M(r), \beta(r)$ and $Q(r),
Q_r(r)$.

\subsection{The mass profile}

We find that we are unable to discriminate among different $M(r)$
models characterized by different inner and asymptotic slopes (see
eqs.~\ref{e:nfw}, \ref{e:ein}, \ref{e:her},
\ref{e:bur}). Independently of the model considered, the concentration
of the mass profile, $c=4.0_{-0.6}^{+1.0}$, is only slightly above
(but not significantly different from) the theoretical expectations by
\citet{BHHV13} and \citet{DeBoni+13} for a cluster of the same mass
and at same redshift as our stacked cluster (see Fig.~\ref{f:cM}). It
is also consistent, albeit slightly below, the observational relation
of \citet{GGS16} which is based on a heterogeneous sample of 293
clusters and derived using different techniques (lensing, X-ray,
kinematics).

\begin{table}
\centering
\caption{Compilation of concentration values}
\label{t:clit}
\begin{tabular}{crccll}
\hline 
\\[-0.2cm]
$z$ range & $\overline{M}_{200}$ & $c,c_{\star}$ & Ref & Note \\[0.11cm]
\hline
\multicolumn{5}{c}{Total mass} \\[0.15cm]
\\[-0.2cm]
0.02-0.13 &     0.7  &    $3.0_{-1.2}^{+3.7}$  &  BG03   & S43 \\[0.11cm]
0.04-0.08 &     6.1  &    $4.0_{-1.5}^{+2.7}$  &  KBM04  & S59 \\[0.11cm]
0.17-0.55 &     8.2  &    $3.6_{-0.4}^{+0.4}$  &  vdM+00 & S16 \\[0.11cm]
0.19-0.35 &    13.1  &    $3.3_{-0.2}^{+0.2}$  &  Mer+15 & A9 \\[0.11cm]
0.36-0.54 &    10.2  &    $4.0_{-0.6}^{+0.6}$  &  Mer+15 & A8 \\[0.11cm]
0.39-0.79 &     2.8  &    $3.2_{-1.0}^{+1.2}$  &  BP09   & S19 \\[0.11cm]
0.69-0.89 &     1.1  &    $4.0_{-0.7}^{+0.7}$  &  Mer+15 & A2 \\[0.11cm]
0.78-1.46 &     4.1  &    $2.3_{-0.2}^{+0.2}$  &  SC13   & S31 \\[0.11cm] 
0.87-1.33 &     4.5  &    $4.0_{-0.6}^{+1.0}$  &  This paper & S10 \\[0.11cm]
\hline
\multicolumn{5}{c}{Stellar mass} \\[0.15cm]
0.01-0.09 &     2.0  &    $2.9_{-0.2}^{+0.2}$  &  LMS04  & S93 \\[0.11cm]
0.04-0.11 &     5.0  &    $1.6_{-0.1}^{+0.5}$  &  vdB+15 & S31 \\[0.11cm]
0.11-0.25 &     5.0  &    $2.1_{-0.2}^{+0.5}$  &  vdB+15 & S30 \\[0.11cm]
0.19-0.55 &    12.0  &    $4.3_{-0.7}^{+0.7}$  &  Muz+07 & S15 \\[0.11cm]
0.44-0.44 &    14.0  &    $3.8_{-0.5}^{+0.5}$  &  Ann+14 & 1 \\[0.11cm]
0.87-1.33 &     4.5  &    $6.7_{-1.1}^{+1.3}$  &  This paper & S10  \\[0.11cm]
\hline
\end{tabular}
\tablefoot{References: BG03: \citet{BG03}, KBM04: \citet{KBM04},
  vdM+00: \citet{vanderMarel+00}, Mer+15: \citet{Merten+15}, BP09:
  \citet{BP09}, SC13: \citet{SC13}, LMS04: \citet{LMS04}, vdB+15:
  \citet{vanderBurg+15}, Muz+07: \citet{Muzzin+07}, Ann+14:
  \citet{Annunziatella+14}. Notes: 'S' indicates the listed value has
  been obtained for a stack sample, 'A' indicates the listed value is
  an average of many, and the following number indicates how many
  clusters were used. }
\end{table}

\begin{figure*}
\begin{center}
\begin{minipage}{0.9\textwidth}
\resizebox{\hsize}{!}{\includegraphics{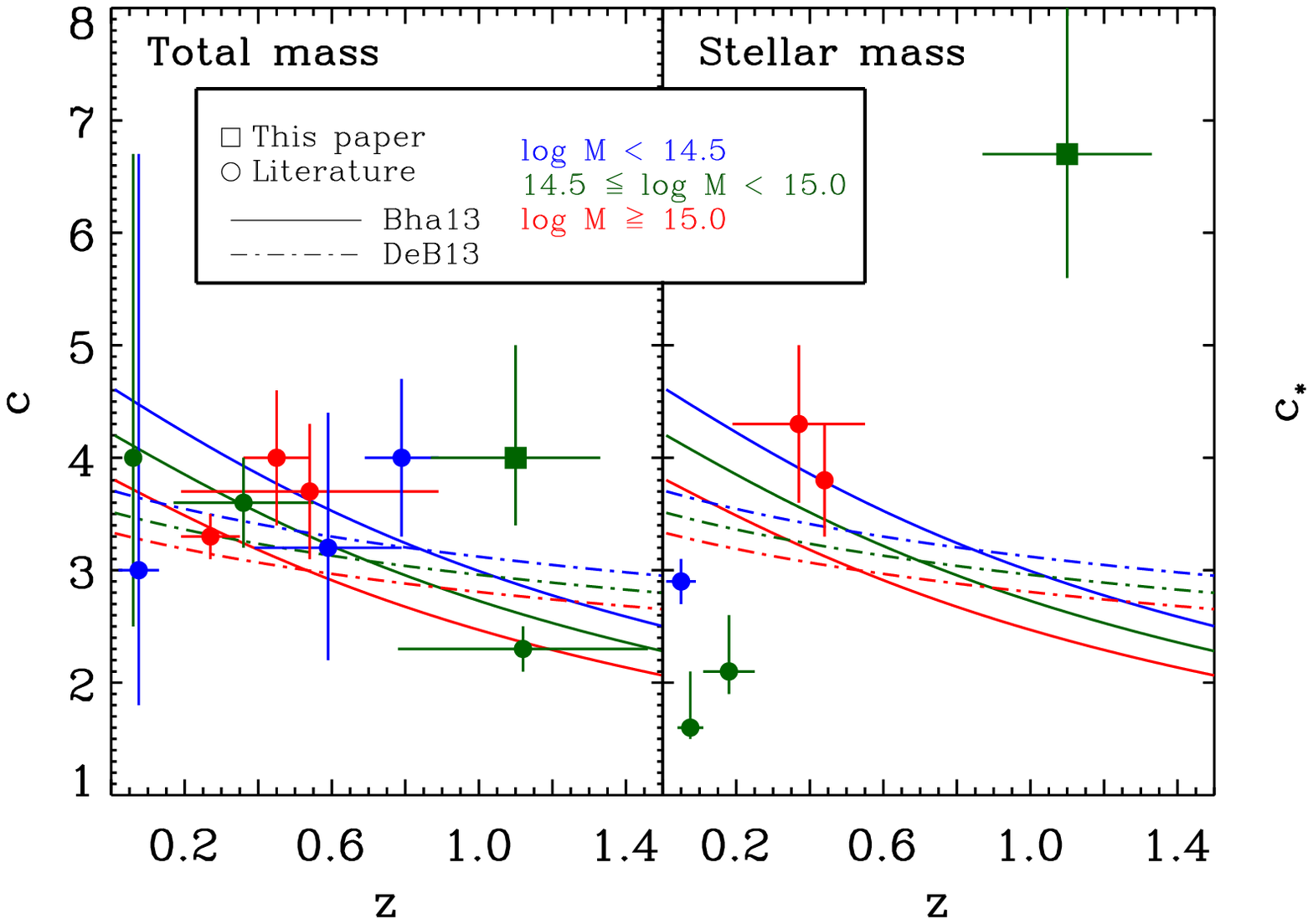}}
\end{minipage}
\end{center}
\caption{The theoretical concentration vs. redshift relations for
  halos of masses $\log \mtwo/\msun=14.25, 14.75, 15.25$ (blue, green,
  red curves, resp.), from \citet{BHHV13} (solid lines) and from
  \citet{DeBoni+13} (dash-dotted lines).  The data-points represent
  determination of the total and stellar mass concentrations (left and
  right panel, resp.)  from the literature and the present work
  (indicated by a square). The points are colored according to the
  average mass of the sample of clusters used for the determination of
  $c$: $\log \mtwo/\msun < 14.5$ (blue), $14.5 \leq \log \mtwo/\msun <
  15.0$ (green), $\log \mtwo/\msun \geq 15.0$ (red). The data
  references can be found in Table~\ref{t:clit}.}
\label{f:cz}
\end{figure*}

The concentration of the total cluster mass is not significantly
different from that of the distribution of cluster galaxies
($c_{\rm{G}}=5.1_{-0.4}^{+0.7}$), while both are significantly less
concentrated than the stellar-mass distribution,
$c_{\star}=6.7_{-1.1}^{+1.4}$.  We determine this value by applying
the procedure of \citetalias{vanderBurg+14} 
using our $\rtwo$ scalings (see
Sect.~\ref{s:mprof}). The value we determine is only slightly
different from that of \citetalias{vanderBurg+14}.

In \citet{vanderBurg+15} the concentration of the stellar mass
distribution was shown to evolve from $7 \pm 1$ in the GCLASS sample
to $2.0 \pm 0.3$ in a $0.04<z<0.26$ sample of clusters with masses
that would make them the likely descendants of the GCLASS clusters in
a hierarchical $\Lambda$CDM cosmology. The low- and intermediate-$z$
data used in \citet{vanderBurg+15} are shown in Fig.~\ref{f:cz}, right
panel, with the addition of one data point from
\citet{Annunziatella+14}.  The $c_{\star}$ values shown in the figure
are reported in Table~\ref{t:clit}. In the same figure we also show
the theoretical concentration vs. $z$ relations for clusters of
different masses, obtained for DM-only cosmological simulations by
\citet{BHHV13} and for hydrodynamical cosmological simulations by
\citet{DeBoni+13}.  Clearly, the evolution observed in $c_{\star}$ is
much stronger, and in the opposite sense, of the evolution expected
for the total mass distribution.

\begin{table}
\centering
\caption{Comparison of observed and theoretical concentrations}
\label{t:chi2}
\begin{tabular}{llrr}
\hline 
\\[-0.2cm]
Concentration & Model & $N$ & $\chi^2$ \\[0.11cm]
\hline
\\[-0.2cm]
total mass ($c$)        & Bha13   & 10 &  12.8 \\[0.11cm]
total mass ($c$)        & DeB13  & 10 &  19.7 \\[0.11cm]
stellar mass ($c_{\star}$) & Bha13   &  6 & 118.1 \\[0.11cm]
stellar mass ($c_{\star}$) & DeB13  &  6 & 51.1 \\[0.11cm]
\hline
\end{tabular}
\tablefoot{References: Bha13: \citet{BHHV13}, DeB13: \citet{DeBoni+13}.
$N$ is the number of data-points used to evaluate $\chi^2$.}
\end{table}

In the left panel of Fig.~\ref{f:cz} we compare the theoretical
$c=c(z)$ expectations to the concentrations derived for the
total mass distributions in clusters of different masses and at
different $z$, using the result of this paper and a compilation of
values from the literature (see Table~\ref{t:clit}). The agreement
is much better than found for the stellar mass concentrations.
We quantify this (dis)agreement by the quantity
\begin{equation}
\chi^2=\sum [(c-c_{\rm{theo}})/\delta c]^2,
\end{equation}
where $c$ are the observed concentrations for the total or stellar
mass, $\delta c$ are their uncertainties, and $c_{\rm{theo}}$ are the
expected concentrations for halos of the same masses and $z$, from
either \citet{BHHV13} or \citet{DeBoni+13}.  The $\chi^2$ values
are given in Table~\ref{t:chi2}. The model of
\citet{BHHV13} is an acceptable description of the observed total mass
concentrations as a function of $z$, at the 17\% confidence level. The
model of \citet{DeBoni+13} is also marginally acceptable (2\%
confidence level), while the $c_{\star}$ observations cannot be
described by any of the two models.

Neither in the total nor in the stellar mass density profile have the
BCGs been considered. As pointed out by \citet{vanderBurg+15}, the
strong observed evolution of $c_{\star}$ could be due to the build-up
of the BCG via merger of central galaxies. \citet{Lidman+12} have
found that BCGs grow on average by a factor $\sim 2$ in stellar mass
between $z=0.9$ and $z=0.2$, mostly via dry mergers. About half of
this growth is due to major mergers \citep{Lidman+13}.  The so-called
Spiderweb galaxy \citep{Miley+06} is a famous high-$z$ example of a
forming BCG.  In the merging process part of the stellar mass may also
get dispersed into the intra-cluster light component \citep[see,
  e.g.][]{Annunziatella+14,Annunziatella+16}, which has also not been
accounted for, in our stellar mass density profile determination.

Galaxies present in the central region of $z \sim 1$ clusters make
the stellar mass density profile very concentrated. As they merge to
form the central BCG, the stellar mass density profile flattens at the
center, if the contribution of the BCG itself is not accounted for,
and if the merged galaxies are not replaced.  The reason why the
merged galaxies are not replaced in lower-$z$ clusters is possibly due
to the increasing timescale for dynamical friction, as the cluster
grows more massive with time \citep{vanderBurg+15}.

\subsection{The pseudo-phase-space density profiles}
Similarly to $M(r)$, also the pseudo-phase-space density profiles
follow closely the theoretical predictions, namely the power-laws of
\citet{DML05}, at any redshift at which they have been estimated so
far \citep{Biviano+13,MBM14}, including the present study.  However,
the $\qr$ and $\qrr$ power-law slopes of our stack cluster are $\sim
8$\% more negative than the theoretically expected ones of
\citet{DML05}.  Since the latter were derived for $z \sim 0$ halos,
the slight difference might hint to an evolution of the
pseudo-phase-space density profiles.

Such an evolution has indeed been predicted by \citet{LC09}. They
relate $\qr$ with the dynamical entropy of the system, $K(r) \equiv
\qr^{-2/3}$, itself regulated by the mass and the mass accretion rate
of the system, $K \propto (M^2/\dot{M})^{2/3}$. Since $\dot{M}$
evolves with $z$, so does the power-law slope of $\qr$, $\alpha$. The
expected evolution over the redshift range from $z=0$ to 1 is only
10\%. \citet{LC09} predict $\alpha=-2.1$ for $\qr$ at $z=1$ (see their
Fig.~6), in perfect agreement with the observed slope of the $\qr$ of
our stack cluster.  Given the large observational uncertainties, such
a perfect agreement must be considered a coincidence, and more data
are needed to really constrain $\alpha$ as a function of $z$.

A note of caution is that we are here comparing profiles obtained for
the DM component in simulations, with profiles obtained using the
total mass density profile ($\rho(r)$) and the total ($\sigma(r)$) or
radial ($\sigma_r(r)$) velocity dispersion profiles of cluster
galaxies. So the comparison of the observed $\qr$ and $\qrr$ with the
theoretically predicted ones is not fully straightforward, and it
could be affected by any velocity bias between the cluster galaxy and
DM particle components \citep[see, e.g.][]{Biviano+06}. As long as
this velocity bias does not depend on the radius, however, only the
normalization of $\qr$ and $\qrr$ would be affected and not the
power-law slope, which is the quantity we are interested in
here. Turning this argument around, the good agreement between the
theoretical and observed slopes suggests that any scale-dependent
velocity bias between the DM and galaxy components of galaxy clusters
must be small, and this is perhaps not too surprising given that they
are both nearly collisionless components.

Overall, the consistency of the $M(r)$ and $\qr$ observed evolution
with the theoretical expectations suggest that our current
understanding of the evolution of the DM component of massive halos is
correct. It indicates that GCLASS $z \sim 1$ clusters are
dynamically evolved systems. Given that no well-formed cluster has so
far been observed at $z>2$ \citep{Newman+14}, this leaves $\sim 2$
Gyr for clusters to reach dynamical equilibrium after their initial
assembly.  Such a short time suggests that the origin of the power-law
behavior of $\qr$ (and its universality) is violent relaxation
\citep{LyndenBell67}.

\subsection{The velocity anisotropy profiles}
The $\br$ of our $z \sim 1$ clusters is very similar to that predicted
for DM particles in cluster-size halos from numerical simulations
\citep[e.g.][]{MBM10,Munari+13}, namely nearly isotropic near the
cluster center and increasingly elongated with radius. This shape is
as predicted for a cosmological halo evolving through an initial phase
of fast collapse and a subsequent slow phase of inside-out growth by
accretion of field material \citep{LC11}.  The $z \sim 1$ cluster
$\br$ is also similar to the $\br$ found for cluster galaxies at
intermediate-$z$ \citep{BP09,Biviano+13,Annunziatella+16}.  Both at $z
\sim 1$ and at intermediate-$z$ PG and SFG have similar $\br$.  Given
that the PG population grows from the quenching of SFG, the observed
similarity of the orbits of these two classes of galaxies at
intermediate- and high-$z$ is not surprising.  At low-$z$, however,
the orbits of SFG and PG become different, as PG develop nearly
isotropic orbits at all radii out to $\sim \rtwo$, while the orbits
of SFG do not change \citep{Mahdavi+99,vanderMarel+00,BK04,KBM04,BP09,MBM14}.

Observations suggest that the evolution of PG orbits happens at
  $z \lesssim 0.2$, i.e. in the last $\lesssim 2$ Gyr of cosmic
  time. Before that, orbital evolution processes might be masked by
  the continuous replenishement of the cluster population by recently
  infallen galaxies.  These are SFG that will have their
  star-formation quenched by the cluster environment, or galaxies that
  are already passive when they enter the cluster
  \citep['pre-processing'; see][]{BNM00}. Only when the mass
  accretion rate declines significantly, and the cluster PG population
  is mostly frozen, its orbital distribution can evolve. The $\sim 2$
  Gyr timescale for orbital evolution then sets an upper limit to the
  quenching timescale, or we would be able to observe a similar
  orbital evolution for the SFG population.

There are several processes that could in principle cause orbital
isotropization, such as dynamical friction, violent relaxation
following major mass accretion by the clusters, modification of the
gravitational potential by secular mass accretion \citep{Gill+04},
radial orbital instability \citep{Bellovary+08}, and interaction with
the intra-cluster medium \citep{DBMS09}. Insight in which of these
processes is more effective in shaping galaxy orbits can come from an
estimate of their relative timescales at different epochs of the
  cluster evolution. 

Numerical simulations can be very useful in this context. So far they
have succeeded in reproducing the overall shape of the $\br$ of
cluster galaxies, {\bf but unfortunately} they have produced
discordant results on its evolution. Some simulations predict marginal
isotropization of the orbits of galaxies (or DM particles) with time
\citep{Wetzel11,Munari+13}, at least in qualitative agreement with the
observed evolution, while others predict no $\br$ evolution at all or
an evolution in the opposite sense \citep{Lemze+12,ID12}. In addition,
\citet{ID12} predict higher $\br$ for PG than for SFG, at variance
with observations.

A better characterization of the orbital evolution as a function of
$z$ and of the cluster internal properties could help us to better
constrain the timescales of galaxy evolution in clusters.

\section{Summary and conclusions}
\label{s:conc}
Using a sample of 10 clusters from the GCLASS survey we build a stack
cluster at $\overline{z}=1.02 \pm 0.06$ and with an average mass
$\mtwo=(4.5 \pm 1.2) \, 10^{14} \, \msun$, that we infer from the
clusters $\slos$ via a scaling relation. The stack cluster contains
418 member galaxies with $z$, 355 within $\rtwo$. We run the
\texttt{MAMPOSSt} algorithm \citep{MBB13} to constrain the
scale-radius of the stack cluster $M(r)$, by considering four
different models for the mass distribution and three different models
for the velocity anisotropy profile $\br$.  Model-independent results
for $\br$ are then obtained from the direct inversion of the Jeans
equation \citep{SSS90}, given the $M(r)$ best-fit obtained via
\texttt{MAMPOSSt}. Using the results of the \texttt{MAMPOSSt} and
Jeans inversion analyses, we finally determine the pseudo-phase-space
density profiles $\qr$ and $\qrr$ for the stack cluster. We compare
our results with those obtained for lower-$z$ clusters to constrain
the evolution of $M(r), \beta(r)$, and $\qr$.  Our results can be
summarized as follows.

\begin{itemize}
\item We constrain the $M(r)$ concentration $c \equiv
  \rtwo/\rs=4.0_{-0.6}^{+1.0}$. This value is in agreement with
  concentration-mass relations derived from cosmological simulations
  \citep{DeBoni+13,BHHV13}, and from observations \citep{GGS16}, and
  significantly smaller than the concentrations of the stellar-mass
  distributions in the same clusters \citepalias{vanderBurg+14}.  The
  evolution $c=c(z)$ agrees with the theoretical expectations from
  cosmological numerical simulations.  The evolution of the
  concentration of the stellar mass distribution
  $c_{\star}=c_{\star}(z)$ is markedly different, and can be explained
  by merging processes of central galaxies leading to the formation of
  the BCG \citep{Lidman+12,Lidman+13,vanderBurg+15}.
\item We find $\br$ to be similar to the $\br$ of DM particles in
  cluster-size halos from cosmological simulations
  \citep[e.g.][]{MBM10,Munari+13} and as expected from the theoretical
  model of \citet{LC11}.  This profile is characterized by isotropic
  orbits near the cluster center, and increasingly radially elongated
  orbits with increasing cluster-centric distance. PG and SFG have
  statistically indistinguishable orbits, similar to those found in
  intermediate-$z$ clusters, and to those of SFG in low-$z$ clusters,
  where the PG have instead isotropic orbits at all radii.  While we
  are unable to identify the physical process responsible for the PG
  orbital evolution, we argue that it must be characterized by a
  longer timescale than the process of SFG quenching (and
  transformation into PG).
\item $\qr$ and $\qrr$ are similar to the theoretically predicted
  power-law relations for cosmologically simulated DM halos of
  \citet{DML05}, with a hint of a slightly steeper slope, in line with
  evolutionary predictions from the theoretical model of \citet{LC09}.
  This indicates that the process leading to the establishment of the
  universal power-law shape of $\qr$ and $\qrr$ must be a rapid one,
  like, e.g., violent relaxation.  Clusters at $z \approx 1$ therefore
  seem to have already attained dynamical equilibrium.
\end{itemize}

With the advent of deeper spectroscopic surveys of cluster galaxies in
the near future at low- \citep[OmegaWINGS,][]{Gullieuszik+15} and
intermediate-$z$ \citep[CLASH-VLT,][]{Rosati+14} it will be possible to
gain a better characterization of the internal cluster dynamics out to
$z \sim 0.5$, and this will help us put in a more constrained context
the results for $z \sim 1$ clusters presented in this work.

\begin{acknowledgements}
We wish to thank R. Capasso, A. Cavaliere, A. Lapi, N. Napolitano, and
A. Saro for useful discussion.  Financial support for this work was
provided by the PRIN INAF 2014: "Glittering kaleidoscopes in the sky:
the multifaceted nature and role of Galaxy Clusters", P.I.: Mario
Nonino, by NSF grant AST-1517863, and by NASA through programs
GO-13306, GO-13677, GO-13747 \& GO-13845 from the Space Telescope
Science Institute, which is operated by AURA, Inc., under NASA
contract NAS 5-26555. RFJvdB acknowledges support from the European
Research Council under FP7 grant number 340519. BS acknowledges a
grant from ``Consorzio per la Fisica - Trieste'' and the financial
support from the University of Trieste through the program
``Finanziamento di Ateneo per progetti di ricerca scientifica - FRA
2015''. GW acknowledges financial support for this work from NSF grant
AST-1517863 and from NASA through programs GO-13306, GO-13677,
GO-13747 and GO-13845/14327 from the Space Telescope Science Institute,
which is operated by AURA, Inc., under NASA contract NAS 5-26555.
\end{acknowledgements}

\bibliography{master}

\begin{thebibliography}{101}
\expandafter\ifx\csname natexlab\endcsname\relax\def\natexlab#1{#1}\fi

\bibitem[{{Adam} {et~al.}(2015){Adam}, {Comis}, {Mac{\'{\i}}as-P{\'e}rez},
  {Adane}, {Ade}, {Andr{\'e}}, {Beelen}, {Belier}, {Beno{\^i}t}, {Bideaud},
  {Billot}, {Blanquer}, {Bourrion}, {Calvo}, {Catalano}, {Coiffard},
  {Cruciani}, {D'Addabbo}, {D{\'e}sert}, {Doyle}, {Goupy}, {Kramer},
  {Leclercq}, {Martino}, {Mauskopf}, {Mayet}, {Monfardini}, {Pajot}, {Pascale},
  {Perotto}, {Pointecouteau}, {Ponthieu}, {Rev{\'e}ret}, {Ritacco},
  {Rodriguez}, {Savini}, {Schuster}, {Sievers}, {Tucker}, \& {Zylka}}]{Adam+15}
{Adam}, R., {Comis}, B., {Mac{\'{\i}}as-P{\'e}rez}, J.-F., {et~al.} 2015, \aap,
  576, A12

\bibitem[{{Annunziatella} {et~al.}(2014){Annunziatella}, {Biviano}, {Mercurio},
  {Nonino}, {Rosati}, {Balestra}, {Presotto}, {Girardi}, {Gobat}, {Grillo},
  {Kelson}, {Medezinski}, {Postman}, {Scodeggio}, {Brescia}, {Demarco},
  {Fritz}, {Koekemoer}, {Lemze}, {Lombardi}, {Sartoris}, {Umetsu}, {Vanzella},
  {Bradley}, {Coe}, {Donahue}, {Infante}, {Kuchner}, {Maier}, {Reg{\H o}s},
  {Verdugo}, \& {Ziegler}}]{Annunziatella+14}
{Annunziatella}, M., {Biviano}, A., {Mercurio}, A., {et~al.} 2014, \aap, 571,
  A80

\bibitem[{{Annunziatella} {et~al.}(2016){Annunziatella}, {Mercurio}, {Biviano},
  {Girardi}, {Nonino}, {Balestra}, {Rosati}, {Bartosch Caminha}, {Brescia},
  {Gobat}, {Grillo}, {Lombardi}, {Sartoris}, {De Lucia}, {Demarco}, {Frye},
  {Fritz}, {Moustakas}, {Scodeggio}, {Kuchner}, {Maier}, \&
  {Ziegler}}]{Annunziatella+16}
{Annunziatella}, M., {Mercurio}, A., {Biviano}, A., {et~al.} 2016, \aap, 585,
  A160

\bibitem[{{Arnaud} {et~al.}(2005){Arnaud}, {Pointecouteau}, \& {Pratt}}]{APP05}
{Arnaud}, M., {Pointecouteau}, E., \& {Pratt}, G.~W. 2005, \aap, 441, 893

\bibitem[{{Balestra} {et~al.}(2016){Balestra}, {Mercurio}, {Sartoris},
  {Girardi}, {Grillo}, {Nonino}, {Rosati}, {Biviano}, {Ettori}, {Forman},
  {Jones}, {Koekemoer}, {Medezinski}, {Merten}, {Ogrean}, {Tozzi}, {Umetsu},
  {Vanzella}, {van Weeren}, {Zitrin}, {Annunziatella}, {Caminha}, {Broadhurst},
  {Coe}, {Donahue}, {Fritz}, {Frye}, {Kelson}, {Lombardi}, {Maier},
  {Meneghetti}, {Monna}, {Postman}, {Scodeggio}, {Seitz}, \&
  {Ziegler}}]{Balestra+16}
{Balestra}, I., {Mercurio}, A., {Sartoris}, B., {et~al.} 2016, \apjs, 224, 33

\bibitem[{{Balogh} {et~al.}(2000){Balogh}, {Navarro}, \& {Morris}}]{BNM00}
{Balogh}, M.~L., {Navarro}, J.~F., \& {Morris}, S.~L. 2000, \apj, 540, 113

\bibitem[{{Beers} {et~al.}(1990){Beers}, {Flynn}, \& {Gebhardt}}]{BFG90}
{Beers}, T.~C., {Flynn}, K., \& {Gebhardt}, K. 1990, \aj, 100, 32

\bibitem[{{Beers} {et~al.}(1991){Beers}, {Gebhardt}, {Forman}, {Huchra}, \&
  {Jones}}]{Beers+91}
{Beers}, T.~C., {Gebhardt}, K., {Forman}, W., {Huchra}, J.~P., \& {Jones}, C.
  1991, \aj, 102, 1581

\bibitem[{{Bellovary} {et~al.}(2008){Bellovary}, {Dalcanton}, {Babul}, {Quinn},
  {Maas}, {Austin}, {Williams}, \& {Barnes}}]{Bellovary+08}
{Bellovary}, J.~M., {Dalcanton}, J.~J., {Babul}, A., {et~al.} 2008, \apj, 685,
  739

\bibitem[{{Bhattacharya} {et~al.}(2013){Bhattacharya}, {Habib}, {Heitmann}, \&
  {Vikhlinin}}]{BHHV13}
{Bhattacharya}, S., {Habib}, S., {Heitmann}, K., \& {Vikhlinin}, A. 2013, \apj,
  766, 32

\bibitem[{{Binney} \& {Mamon}(1982)}]{BM82}
{Binney}, J. \& {Mamon}, G.~A. 1982, \mnras, 200, 361

\bibitem[{{Binney} \& {Tremaine}(1987)}]{BT87}
{Binney}, J. \& {Tremaine}, S. 1987, Galactic dynamics (Princeton, NJ,
  Princeton University Press, 1987, 747 p.)

\bibitem[{{Biviano} \& {Girardi}(2003)}]{BG03}
{Biviano}, A. \& {Girardi}, M. 2003, \apj, 585, 205

\bibitem[{{Biviano} \& {Katgert}(2004)}]{BK04}
{Biviano}, A. \& {Katgert}, P. 2004, \aap, 424, 779

\bibitem[{{Biviano} {et~al.}(1997){Biviano}, {Katgert}, {Mazure}, {Moles}, {den
  Hartog}, {Perea}, \& {Focardi}}]{Biviano+97}
{Biviano}, A., {Katgert}, P., {Mazure}, A., {et~al.} 1997, \aap, 321, 84

\bibitem[{{Biviano} {et~al.}(2006){Biviano}, {Murante}, {Borgani}, {Diaferio},
  {Dolag}, \& {Girardi}}]{Biviano+06}
{Biviano}, A., {Murante}, G., {Borgani}, S., {et~al.} 2006, \aap, 456, 23

\bibitem[{{Biviano} \& {Poggianti}(2009)}]{BP09}
{Biviano}, A. \& {Poggianti}, B.~M. 2009, \aap, 501, 419

\bibitem[{{Biviano} {et~al.}(2013){Biviano}, {Rosati}, {Balestra}, {Mercurio},
  {Girardi}, {Nonino}, {Grillo}, {Scodeggio}, {Lemze}, {Kelson}, {Umetsu},
  {Postman}, {Zitrin}, {Czoske}, {Ettori}, {Fritz}, {Lombardi}, {Maier},
  {Medezinski}, {Mei}, {Presotto}, {Strazzullo}, {Tozzi}, {Ziegler},
  {Annunziatella}, {Bartelmann}, {Benitez}, {Bradley}, {Brescia}, {Broadhurst},
  {Coe}, {Demarco}, {Donahue}, {Ford}, {Gobat}, {Graves}, {Koekemoer},
  {Kuchner}, {Melchior}, {Meneghetti}, {Merten}, {Moustakas}, {Munari}, {Reg{\H
  o}s}, {Sartoris}, {Seitz}, \& {Zheng}}]{Biviano+13}
{Biviano}, A., {Rosati}, P., {Balestra}, I., {et~al.} 2013, \aap, 558, A1

\bibitem[{{Burkert}(1995)}]{Burkert95}
{Burkert}, A. 1995, \apjl, 447, L25

\bibitem[{{Carlberg} {et~al.}(1997{\natexlab{a}}){Carlberg}, {Yee},
  {Ellingson}, {Morris}, {Abraham}, {Gravel}, {Pritchet}, {Smecker-Hane},
  {Hartwick}, {Hesser}, {Hutchings}, \& {Oke}}]{Carlberg+97-mprof}
{Carlberg}, R.~G., {Yee}, H. K.~C., {Ellingson}, E., {et~al.}
  1997{\natexlab{a}}, \apjl, 485, L13

\bibitem[{{Carlberg} {et~al.}(1997{\natexlab{b}}){Carlberg}, {Yee},
  {Ellingson}, {Morris}, {Abraham}, {Gravel}, {Pritchet}, {Smecker-Hane},
  {Hartwick}, {Hesser}, {Hutchings}, \& {Oke}}]{Carlberg+97-equil}
{Carlberg}, R.~G., {Yee}, H. K.~C., {Ellingson}, E., {et~al.}
  1997{\natexlab{b}}, \apjl, 476, L7

\bibitem[{{Cava} {et~al.}(2009){Cava}, {Bettoni}, {Poggianti}, {Couch},
  {Moles}, {Varela}, {Biviano}, {D'Onofrio}, {Dressler}, {Fasano}, {Fritz},
  {Kj{\ae}rgaard}, {Ramella}, \& {Valentinuzzi}}]{Cava+09}
{Cava}, A., {Bettoni}, D., {Poggianti}, B.~M., {et~al.} 2009, \aap, 495, 707

\bibitem[{{Clowe} {et~al.}(2000){Clowe}, {Luppino}, {Kaiser}, \&
  {Gioia}}]{Clowe+00}
{Clowe}, D., {Luppino}, G.~A., {Kaiser}, N., \& {Gioia}, I.~M. 2000, \apj, 539,
  540

\bibitem[{{Danese} {et~al.}(1980){Danese}, {de Zotti}, \& {di
  Tullio}}]{DdZdT80}
{Danese}, L., {de Zotti}, G., \& {di Tullio}, G. 1980, \aap, 82, 322

\bibitem[{{De Boni} {et~al.}(2013){De Boni}, {Ettori}, {Dolag}, \&
  {Moscardini}}]{DeBoni+13}
{De Boni}, C., {Ettori}, S., {Dolag}, K., \& {Moscardini}, L. 2013, \mnras,
  428, 2921

\bibitem[{{Dehnen} \& {McLaughlin}(2005)}]{DML05}
{Dehnen}, W. \& {McLaughlin}, D.~E. 2005, \mnras, 363, 1057

\bibitem[{{Dejonghe} \& {Merritt}(1992)}]{DM92}
{Dejonghe}, H. \& {Merritt}, D. 1992, \apj, 391, 531

\bibitem[{{Del Popolo}(2011)}]{DelPopolo11}
{Del Popolo}, A. 2011, \jcap, 7, 014

\bibitem[{{den Hartog} \& {Katgert}(1996)}]{dHK96}
{den Hartog}, R. \& {Katgert}, P. 1996, \mnras, 279, 349

\bibitem[{{Diemer} \& {Kravtsov}(2014)}]{DK14}
{Diemer}, B. \& {Kravtsov}, A.~V. 2014, \apj, 789, 1

\bibitem[{{Dolag} {et~al.}(2009){Dolag}, {Borgani}, {Murante}, \&
  {Springel}}]{DBMS09}
{Dolag}, K., {Borgani}, S., {Murante}, G., \& {Springel}, V. 2009, \mnras, 399,
  497

\bibitem[{{Dressler} \& {Shectman}(1988)}]{DS88}
{Dressler}, A. \& {Shectman}, S.~A. 1988, \aj, 95, 985

\bibitem[{{Durret} {et~al.}(2015){Durret}, {Wakamatsu}, {Nagayama}, {Adami}, \&
  {Biviano}}]{Durret+15}
{Durret}, F., {Wakamatsu}, K., {Nagayama}, T., {Adami}, C., \& {Biviano}, A.
  2015, \aap, 583, A124

\bibitem[{{Einasto}(1965)}]{Einasto65}
{Einasto}, J. 1965, Trudy Astrofizicheskogo Instituta Alma-Ata, 5, 87

\bibitem[{{El-Zant}(2008)}]{ElZant08}
{El-Zant}, A.~A. 2008, \apj, 681, 1058

\bibitem[{{Fadda} {et~al.}(1996){Fadda}, {Girardi}, {Giuricin}, {Mardirossian},
  \& {Mezzetti}}]{Fadda+96}
{Fadda}, D., {Girardi}, M., {Giuricin}, G., {Mardirossian}, F., \& {Mezzetti},
  M. 1996, \apj, 473, 670

\bibitem[{{Gao} {et~al.}(2008){Gao}, {Navarro}, {Cole}, {Frenk}, {White},
  {Springel}, {Jenkins}, \& {Neto}}]{Gao+08}
{Gao}, L., {Navarro}, J.~F., {Cole}, S., {et~al.} 2008, \mnras, 387, 536

\bibitem[{{Gebhardt} {et~al.}(1994){Gebhardt}, {Pryor}, {Williams}, \&
  {Hesser}}]{Gebhardt+94}
{Gebhardt}, K., {Pryor}, C., {Williams}, T.~B., \& {Hesser}, J.~E. 1994, \aj,
  107, 2067

\bibitem[{{Gill} {et~al.}(2004){Gill}, {Knebe}, {Gibson}, \&
  {Dopita}}]{Gill+04}
{Gill}, S.~P.~D., {Knebe}, A., {Gibson}, B.~K., \& {Dopita}, M.~A. 2004,
  \mnras, 351, 410

\bibitem[{{Girardi} {et~al.}(2011){Girardi}, {Bardelli}, {Barrena}, {Boschin},
  {Gastaldello}, \& {Nonino}}]{Girardi+11}
{Girardi}, M., {Bardelli}, S., {Barrena}, R., {et~al.} 2011, \aap, 536, A89

\bibitem[{{Gladders} \& {Yee}(2000)}]{GY00}
{Gladders}, M.~D. \& {Yee}, H.~K.~C. 2000, \aj, 120, 2148

\bibitem[{{Groener} {et~al.}(2016){Groener}, {Goldberg}, \& {Sereno}}]{GGS16}
{Groener}, A.~M., {Goldberg}, D.~M., \& {Sereno}, M. 2016, \mnras, 455, 892

\bibitem[{{Guennou} {et~al.}(2014){Guennou}, {Biviano}, {Adami}, {Limousin},
  {Lima Neto}, {Mamon}, {Ulmer}, {Gavazzi}, {Cypriano}, {Durret}, {Clowe},
  {LeBrun}, {Allam}, {Basa}, {Benoist}, {Cappi}, {Halliday}, {Ilbert},
  {Johnston}, {Jullo}, {Just}, {Kubo}, {M{\'a}rquez}, {Marshall}, {Martinet},
  {Maurogordato}, {Mazure}, {Murphy}, {Plana}, {Rostagni}, {Russeil},
  {Schirmer}, {Schrabback}, {Slezak}, {Tucker}, {Zaritsky}, \&
  {Ziegler}}]{Guennou+14}
{Guennou}, L., {Biviano}, A., {Adami}, C., {et~al.} 2014, \aap, 566, A149

\bibitem[{{Gullieuszik} {et~al.}(2015){Gullieuszik}, {Poggianti}, {Fasano},
  {Zaggia}, {Paccagnella}, {Moretti}, {Bettoni}, {D'Onofrio}, {Couch},
  {Vulcani}, {Fritz}, {Omizzolo}, {Baruffolo}, {Schipani}, {Capaccioli}, \&
  {Varela}}]{Gullieuszik+15}
{Gullieuszik}, M., {Poggianti}, B., {Fasano}, G., {et~al.} 2015, \aap, 581, A41

\bibitem[{{Hansen}(2009)}]{Hansen09}
{Hansen}, S.~H. 2009, \apj, 694, 1250

\bibitem[{{H{\'e}non}(1964)}]{Henon64}
{H{\'e}non}, M. 1964, Annales d'Astrophysique, 27, 83

\bibitem[{{Hernquist}(1990)}]{Hernquist90}
{Hernquist}, L. 1990, \apj, 356, 359

\bibitem[{{Huo} {et~al.}(2004){Huo}, {Xue}, {Xu}, {Squires}, \&
  {Rosati}}]{Huo+04}
{Huo}, Z.-Y., {Xue}, S.-J., {Xu}, H., {Squires}, G., \& {Rosati}, P. 2004, \aj,
  127, 1263

\bibitem[{{Huss} {et~al.}(1999){Huss}, {Jain}, \& {Steinmetz}}]{HJS99_MN}
{Huss}, A., {Jain}, B., \& {Steinmetz}, M. 1999, \mnras, 308, 1011

\bibitem[{{Hwang} \& {Lee}(2008)}]{HL08}
{Hwang}, H.~S. \& {Lee}, M.~G. 2008, \apj, 676, 218

\bibitem[{{Iannuzzi} \& {Dolag}(2012)}]{ID12}
{Iannuzzi}, F. \& {Dolag}, K. 2012, \mnras, 427, 1024

\bibitem[{{Jee} {et~al.}(2014){Jee}, {Hughes}, {Menanteau}, {Sif{\'o}n},
  {Mandelbaum}, {Barrientos}, {Infante}, \& {Ng}}]{Jee+14}
{Jee}, M.~J., {Hughes}, J.~P., {Menanteau}, F., {et~al.} 2014, \apj, 785, 20

\bibitem[{{Jee} {et~al.}(2009){Jee}, {Rosati}, {Ford}, {Dawson}, {Lidman},
  {Perlmutter}, {Demarco}, {Strazzullo}, {Mullis}, {B{\"o}hringer}, \&
  {Fassbender}}]{Jee+09}
{Jee}, M.~J., {Rosati}, P., {Ford}, H.~C., {et~al.} 2009, \apj, 704, 672

\bibitem[{{Jee} \& {Tyson}(2009)}]{JT09}
{Jee}, M.~J. \& {Tyson}, J.~A. 2009, \apj, 691, 1337

\bibitem[{{Jee} {et~al.}(2005{\natexlab{a}}){Jee}, {White}, {Ben{\'{\i}}tez},
  {Ford}, {Blakeslee}, {Rosati}, {Demarco}, \& {Illingworth}}]{Jee+05}
{Jee}, M.~J., {White}, R.~L., {Ben{\'{\i}}tez}, N., {et~al.}
  2005{\natexlab{a}}, \apj, 618, 46

\bibitem[{{Jee} {et~al.}(2005{\natexlab{b}}){Jee}, {White}, {Ford},
  {Blakeslee}, {Illingworth}, {Coe}, \& {Tran}}]{Jee+05b}
{Jee}, M.~J., {White}, R.~L., {Ford}, H.~C., {et~al.} 2005{\natexlab{b}}, \apj,
  634, 813

\bibitem[{{Jee} {et~al.}(2006){Jee}, {White}, {Ford}, {Illingworth},
  {Blakeslee}, {Holden}, \& {Mei}}]{Jee+06}
{Jee}, M.~J., {White}, R.~L., {Ford}, H.~C., {et~al.} 2006, \apj, 642, 720

\bibitem[{{Katgert} {et~al.}(2004){Katgert}, {Biviano}, \& {Mazure}}]{KBM04}
{Katgert}, P., {Biviano}, A., \& {Mazure}, A. 2004, \apj, 600, 657

\bibitem[{{Kneib} {et~al.}(2003){Kneib}, {Hudelot}, {Ellis}, {Treu}, {Smith},
  {Marshall}, {Czoske}, {Smail}, \& {Natarajan}}]{Kneib+03}
{Kneib}, J.-P., {Hudelot}, P., {Ellis}, R.~S., {et~al.} 2003, \apj, 598, 804

\bibitem[{{Knollmann} {et~al.}(2008){Knollmann}, {Knebe}, \& {Hoffman}}]{KKH08}
{Knollmann}, S.~R., {Knebe}, A., \& {Hoffman}, Y. 2008, \mnras, 391, 559

\bibitem[{{Lapi} \& {Cavaliere}(2009)}]{LC09}
{Lapi}, A. \& {Cavaliere}, A. 2009, \apj, 692, 174

\bibitem[{{Lapi} \& {Cavaliere}(2011)}]{LC11}
{Lapi}, A. \& {Cavaliere}, A. 2011, \apj, 743, 127

\bibitem[{{Lemze} {et~al.}(2012){Lemze}, {Wagner}, {Rephaeli}, {Sadeh},
  {Norman}, {Barkana}, {Broadhurst}, {Ford}, \& {Postman}}]{Lemze+12}
{Lemze}, D., {Wagner}, R., {Rephaeli}, Y., {et~al.} 2012, \apj, 752, 141

\bibitem[{{Lidman} {et~al.}(2013){Lidman}, {Iacobuta}, {Bauer}, {Barrientos},
  {Cerulo}, {Couch}, {Delaye}, {Demarco}, {Ellingson}, {Faloon}, {Gilbank},
  {Huertas-Company}, {Mei}, {Meyers}, {Muzzin}, {Noble}, {Nantais}, {Rettura},
  {Rosati}, {S{\'a}nchez-Janssen}, {Strazzullo}, {Webb}, {Wilson}, {Yan}, \&
  {Yee}}]{Lidman+13}
{Lidman}, C., {Iacobuta}, G., {Bauer}, A.~E., {et~al.} 2013, \mnras, 433, 825

\bibitem[{{Lidman} {et~al.}(2012){Lidman}, {Suherli}, {Muzzin}, {Wilson},
  {Demarco}, {Brough}, {Rettura}, {Cox}, {DeGroot}, {Yee}, {Gilbank},
  {Hoekstra}, {Balogh}, {Ellingson}, {Hicks}, {Nantais}, {Noble}, {Lacy},
  {Surace}, \& {Webb}}]{Lidman+12}
{Lidman}, C., {Suherli}, J., {Muzzin}, A., {et~al.} 2012, \mnras, 427, 550

\bibitem[{{Lin} {et~al.}(2004){Lin}, {Mohr}, \& {Stanford}}]{LMS04}
{Lin}, Y.-T., {Mohr}, J.~J., \& {Stanford}, S.~A. 2004, \apj, 610, 745

\bibitem[{{Lombardi} {et~al.}(2005){Lombardi}, {Rosati}, {Blakeslee}, {Ettori},
  {Demarco}, {Ford}, {Illingworth}, {Clampin}, {Hartig}, {Ben{\'{\i}}tez},
  {Broadhurst}, {Franx}, {Jee}, {Postman}, \& {White}}]{Lombardi+05}
{Lombardi}, M., {Rosati}, P., {Blakeslee}, J.~P., {et~al.} 2005, \apj, 623, 42

\bibitem[{{Lynden-Bell}(1967)}]{LyndenBell67}
{Lynden-Bell}, D. 1967, \mnras, 136, 101

\bibitem[{{Macci{\`o}} {et~al.}(2008){Macci{\`o}}, {Dutton}, \& {van den
  Bosch}}]{MDvdB08}
{Macci{\`o}}, A.~V., {Dutton}, A.~A., \& {van den Bosch}, F.~C. 2008, \mnras,
  391, 1940

\bibitem[{{Mahdavi} {et~al.}(1999){Mahdavi}, {Geller}, {B{\"o}hringer},
  {Kurtz}, \& {Ramella}}]{Mahdavi+99}
{Mahdavi}, A., {Geller}, M.~J., {B{\"o}hringer}, H., {Kurtz}, M.~J., \&
  {Ramella}, M. 1999, \apj, 518, 69

\bibitem[{{Mamon} {et~al.}(2013){Mamon}, {Biviano}, \& {Bou{\'e}}}]{MBB13}
{Mamon}, G.~A., {Biviano}, A., \& {Bou{\'e}}, G. 2013, \mnras, 429, 3079

\bibitem[{{Mamon} {et~al.}(2010){Mamon}, {Biviano}, \& {Murante}}]{MBM10}
{Mamon}, G.~A., {Biviano}, A., \& {Murante}, G. 2010, \aap, 520, A30

\bibitem[{{Mamon} \& {{\L}okas}(2005)}]{ML05b}
{Mamon}, G.~A. \& {{\L}okas}, E.~L. 2005, \mnras, 363, 705

\bibitem[{{Merten} {et~al.}(2015){Merten}, {Meneghetti}, {Postman}, {Umetsu},
  {Zitrin}, {Medezinski}, {Nonino}, {Koekemoer}, {Melchior}, {Gruen},
  {Moustakas}, {Bartelmann}, {Host}, {Donahue}, {Coe}, {Molino}, {Jouvel},
  {Monna}, {Seitz}, {Czakon}, {Lemze}, {Sayers}, {Balestra}, {Rosati},
  {Ben{\'{\i}}tez}, {Biviano}, {Bouwens}, {Bradley}, {Broadhurst}, {Carrasco},
  {Ford}, {Grillo}, {Infante}, {Kelson}, {Lahav}, {Massey}, {Moustakas},
  {Rasia}, {Rhodes}, {Vega}, \& {Zheng}}]{Merten+15}
{Merten}, J., {Meneghetti}, M., {Postman}, M., {et~al.} 2015, \apj, 806, 4

\bibitem[{{Miley} {et~al.}(2006){Miley}, {Overzier}, {Zirm}, {Ford}, {Kurk},
  {Pentericci}, {Blakeslee}, {Franx}, {Illingworth}, {Postman}, {Rosati},
  {R{\"o}ttgering}, {Venemans}, \& {Helder}}]{Miley+06}
{Miley}, G.~K., {Overzier}, R.~A., {Zirm}, A.~W., {et~al.} 2006, \apjl, 650,
  L29

\bibitem[{{Munari} {et~al.}(2013){Munari}, {Biviano}, {Borgani}, {Murante}, \&
  {Fabjan}}]{Munari+13}
{Munari}, E., {Biviano}, A., {Borgani}, S., {Murante}, G., \& {Fabjan}, D.
  2013, \mnras, 430, 2638

\bibitem[{{Munari} {et~al.}(2014){Munari}, {Biviano}, \& {Mamon}}]{MBM14}
{Munari}, E., {Biviano}, A., \& {Mamon}, G.~A. 2014, \aap, 566, A68

\bibitem[{{Muzzin} {et~al.}(2014){Muzzin}, {van der Burg}, {McGee}, {Balogh},
  {Franx}, {Hoekstra}, {Hudson}, {Noble}, {Taranu}, {Webb}, {Wilson}, \&
  {Yee}}]{Muzzin+14}
{Muzzin}, A., {van der Burg}, R.~F.~J., {McGee}, S.~L., {et~al.} 2014, \apj,
  796, 65

\bibitem[{{Muzzin} {et~al.}(2012){Muzzin}, {Wilson}, {Yee}, {Gilbank},
  {Hoekstra}, {Demarco}, {Balogh}, {van Dokkum}, {Franx}, {Ellingson}, {Hicks},
  {Nantais}, {Noble}, {Lacy}, {Lidman}, {Rettura}, {Surace}, \&
  {Webb}}]{Muzzin+12}
{Muzzin}, A., {Wilson}, G., {Yee}, H.~K.~C., {et~al.} 2012, \apj, 746, 188

\bibitem[{{Muzzin} {et~al.}(2009){Muzzin}, {Wilson}, {Yee}, {Hoekstra},
  {Gilbank}, {Surace}, {Lacy}, {Blindert}, {Majumdar}, {Demarco}, {Gardner},
  {Gladders}, \& {Lonsdale}}]{Muzzin+09}
{Muzzin}, A., {Wilson}, G., {Yee}, H.~K.~C., {et~al.} 2009, \apj, 698, 1934

\bibitem[{{Muzzin} {et~al.}(2007){Muzzin}, {Yee}, {Hall}, {Ellingson}, \&
  {Lin}}]{Muzzin+07}
{Muzzin}, A., {Yee}, H.~K.~C., {Hall}, P.~B., {Ellingson}, E., \& {Lin}, H.
  2007, \apj, 659, 1106

\bibitem[{{Navarro} {et~al.}(1996){Navarro}, {Frenk}, \& {White}}]{NFW96}
{Navarro}, J.~F., {Frenk}, C.~S., \& {White}, S. D.~M. 1996, \apj, 462, 563

\bibitem[{{Navarro} {et~al.}(1997){Navarro}, {Frenk}, \& {White}}]{NFW97}
{Navarro}, J.~F., {Frenk}, C.~S., \& {White}, S. D.~M. 1997, \apj, 490, 493

\bibitem[{{Navarro} {et~al.}(2004){Navarro}, {Hayashi}, {Power}, {Jenkins},
  {Frenk}, {White}, {Springel}, {Stadel}, \& {Quinn}}]{Navarro+04}
{Navarro}, J.~F., {Hayashi}, E., {Power}, C., {et~al.} 2004, \mnras, 349, 1039

\bibitem[{{Newman} {et~al.}(2014){Newman}, {Ellis}, {Andreon}, {Treu},
  {Raichoor}, \& {Trinchieri}}]{Newman+14}
{Newman}, A.~B., {Ellis}, R.~S., {Andreon}, S., {et~al.} 2014, \apj, 788, 51

\bibitem[{{Pizzuti} {et~al.}(2016){Pizzuti}, {Sartoris}, {Borgani}, {Amendola},
  {Umetsu}, {Biviano}, {Girardi}, {Rosati}, {Balestra}, {Caminha}, {Frye},
  {Koekemoer}, {Grillo}, {Lombardi}, {Mercurio}, \& {Nonino}}]{Pizzuti+16}
{Pizzuti}, L., {Sartoris}, B., {Borgani}, S., {et~al.} 2016, \jcap, 4, 023

\bibitem[{{Press} {et~al.}(1992){Press}, {Teukolsky}, {Vetterling}, \&
  {Flannery}}]{Press+92}
{Press}, W.~H., {Teukolsky}, S.~A., {Vetterling}, W.~T., \& {Flannery}, B.~P.
  1992, Numerical Recipes in C, 2nd edn. (Cambridge University Press)

\bibitem[{{Rines} {et~al.}(2003){Rines}, {Geller}, {Kurtz}, \&
  {Diaferio}}]{Rines+03}
{Rines}, K., {Geller}, M.~J., {Kurtz}, M.~J., \& {Diaferio}, A. 2003, \aj, 126,
  2152

\bibitem[{{Rosati} {et~al.}(2014){Rosati}, {Balestra}, {Grillo}, {Mercurio},
  {Nonino}, {Biviano}, {Girardi}, {Vanzella}, \& {Clash-VLT Team}}]{Rosati+14}
{Rosati}, P., {Balestra}, I., {Grillo}, C., {et~al.} 2014, The Messenger, 158,
  48

\bibitem[{{Santos} {et~al.}(2012){Santos}, {Tozzi}, {Rosati}, {Nonino}, \&
  {Giovannini}}]{Santos+12}
{Santos}, J.~S., {Tozzi}, P., {Rosati}, P., {Nonino}, M., \& {Giovannini}, G.
  2012, \aap, 539, A105

\bibitem[{{Schaller} {et~al.}(2015){Schaller}, {Frenk}, {Bower}, {Theuns},
  {Trayford}, {Crain}, {Furlong}, {Schaye}, {Dalla Vecchia}, \&
  {McCarthy}}]{Schaller+15}
{Schaller}, M., {Frenk}, C.~S., {Bower}, R.~G., {et~al.} 2015, \mnras, 452, 343

\bibitem[{{Sereno} \& {Covone}(2013)}]{SC13}
{Sereno}, M. \& {Covone}, G. 2013, \mnras, 434, 878

\bibitem[{{Solanes} \& {Salvador-Sol\'e}(1990)}]{SSS90}
{Solanes}, J.~M. \& {Salvador-Sol\'e}, E. 1990, \aap, 234, 93

\bibitem[{{Taylor} \& {Navarro}(2001)}]{TN01}
{Taylor}, J.~E. \& {Navarro}, J.~F. 2001, \apj, 563, 483

\bibitem[{{Tiret} {et~al.}(2007){Tiret}, {Combes}, {Angus}, {Famaey}, \&
  {Zhao}}]{Tiret+07}
{Tiret}, O., {Combes}, F., {Angus}, G.~W., {Famaey}, B., \& {Zhao}, H.~S. 2007,
  \aap, 476, L1

\bibitem[{{van der Burg} {et~al.}(2015){van der Burg}, {Hoekstra}, {Muzzin},
  {Sif{\'o}n}, {Balogh}, \& {McGee}}]{vanderBurg+15}
{van der Burg}, R.~F.~J., {Hoekstra}, H., {Muzzin}, A., {et~al.} 2015, \aap,
  577, A19

\bibitem[{{van der Burg} {et~al.}(2013){van der Burg}, {Muzzin}, {Hoekstra},
  {Lidman}, {Rettura}, {Wilson}, {Yee}, {Hildebrandt}, {Marchesini},
  {Stefanon}, {Demarco}, \& {Kuijken}}]{vanderBurg+13}
{van der Burg}, R.~F.~J., {Muzzin}, A., {Hoekstra}, H., {et~al.} 2013, \aap,
  557, A15

\bibitem[{{van der Burg} {et~al.}(2014){van der Burg}, {Muzzin}, {Hoekstra},
  {Wilson}, {Lidman}, \& {Yee}}]{vanderBurg+14}
{van der Burg}, R.~F.~J., {Muzzin}, A., {Hoekstra}, H., {et~al.} 2014, \aap,
  561, A79

\bibitem[{{van der Marel} {et~al.}(2000){van der Marel}, {Magorrian},
  {Carlberg}, {Yee}, \& {Ellingson}}]{vanderMarel+00}
{van der Marel}, R.~P., {Magorrian}, J., {Carlberg}, R.~G., {Yee}, H.~K.~C., \&
  {Ellingson}, E. 2000, \aj, 119, 2038

\bibitem[{{Wetzel}(2011)}]{Wetzel11}
{Wetzel}, A.~R. 2011, \mnras, 412, 49

\bibitem[{{Wilson} {et~al.}(2009){Wilson}, {Muzzin}, {Yee}, {Lacy}, {Surace},
  {Gilbank}, {Blindert}, {Hoekstra}, {Majumdar}, {Demarco}, {Gardner},
  {Gladders}, \& {Lonsdale}}]{Wilson+09}
{Wilson}, G., {Muzzin}, A., {Yee}, H.~K.~C., {et~al.} 2009, \apj, 698, 1943

\end{thebibliography}

\appendix

\section{Sphericity and dynamical equilibrium}
Our results are based on techniques (\texttt{MAMPOSSt} and the Jeans
inversion technique) that assume spherical symmetry. We do not expect
this assumption to bias our results in a significant way for the
following reasons. GCLASS clusters were chosen from the SpARCS parent
sample based on richnesses measured in apertures of 500~kpc
radius. Since this radius is significantly larger than the typical
scale radius, the cluster orientation should not influence the
selection process. The GCLASS sample of clusters is therefore expected
to be unbiased for a preferential orientation along the line-of-sight.
Stacking aspherical clusters with random orientation is likely to
produce a stack cluster whose dynamics can be adequately investigated
using spherical models, as shown by \citet{vanderMarel+00}.  In
addition, the \texttt{MAMPOSSt} technique has been successfully tested
by \citet{MBB13} on aspherical cluster-sized halos extracted from
cosmological simulations.

Dynamical relaxation is another assumption of the adopted techniques
in this paper. We searched for the presence of substructures in the 10
GCLASS clusters using the technique of \citet{DS88}, and found no
significant evidence for it in any of them (at the 1\% confidence
level). However, the number of members is generally too small in each
individual cluster for this test to be effective \citep[$\gtrsim 50$
  members are generally considered to be necessary,
  see][]{Biviano+97}, with the exception of SpARCS~1613. The test
cannot be run on the stack cluster because the stacking procedure
mixes the projected phase-space distributions of the stacked clusters,
destroying existing correlations of velocities among neighboring
galaxies.

An independent indication that the 10 GCLASS clusters are not far from
dynamical equilibrium is provided by the comparison between the
dynamical masses of the clusters obtained from $\slos$ and their total
stellar masses \citepalias[see Fig.~5 in][]{vanderBurg+14}. Deviation
from dynamical relaxation can boost the estimate of a cluster $\slos$
because of the relative velocities of colliding groups with respect to
the main cluster \citep[e.g.][]{Girardi+11}. Even a small group can
produce a significant change in the estimate of $\slos$, and therefore
of the dynamical mass of the system. On the other hand, this same
group would not contribute much to the total stellar mass estimate of
the cluster. The good correlation existing between the GCLASS cluster
dynamical and stellar masses is therefore suggesting that, overall,
GCLASS clusters are not severely contaminated by infalling groups, and
hence they are not, on average, too far from dynamical equilibrium.

Finally, we note that the validation of the \texttt{MAMPOSSt}
technique was based on a sample of cluster-sized halos which were {\em
  not} selected to be in an advanced stage of dynamical relaxation
\citep{MBB13}.

\end{document}